\documentclass[myadp,fleqn]{w-art}
\usepackage{times}
\usepackage{w-thm}
\usepackage{amsmath}
\usepackage[]{graphicx}
\chardef\bslash=`\\ 

\begin{document}
\newcommand{\be}{\begin{equation}}
\newcommand{\ba}{\begin{eqnarray}}
\newcommand{\ea}{\end{eqnarray}}
\newcommand{\eref}[1]{(\ref{#1})}
\def\C{\mathbf{C}}
\def\R{\mathbf{R}}
\def\Z{\mathbf{Z}}
\def\N{\mathbf{N}}
\renewcommand{\v}[1]{{\bf #1}}
\def\tr{\mathop{\rm Tr}}
\def\mod{\mathop{\rm mod}}
\def\arsinh{\mathop{\rm arsinh}}
\def\arccot{\mathop{\rm arccot}}
\def\up{\uparrow}
\def\down{\downarrow}
\newcommand{\ud}{\mathrm{d}}
\def\Min{\mathop{\rm Min}}
\def\Max{\mathop{\rm Max}}
\def\P{\mathop{\rm P}}
\def\sign{\mathrm{sign}}

\def\Map#1{\smash{\mathop{\hbox to 35 pt{$\,$\rightarrowfill$\,\,$}}
                             \limits^{\scriptstyle#1}}}
\def\MapRight#1{\smash{\mathop{\hbox to 35pt{\rightarrowfill}}\limits^{#1}}}

\def\MapRightLong#1#2#3{\smash{\mathop{\hbox to #1mm{\rightarrowfill}}
   \limits^{\scriptstyle#2}_{\scriptstyle#3}}}

\newcommand{\Rahmen}[1]{\vspace{0.2cm}\par%
   \setlength{\dimen0}{\textwidth}
   \addtolength{\dimen0}{-0.3cm}
   \hspace{-0.6cm}\fbox{\parbox{\dimen0}{#1}}\vspace{0.2cm}\par}

\DOIsuffix{theDOIsuffix}
\Volume{}
\Issue{}
\Copyrightissue{}
\Month{}
\Year{2003}
\pagespan{1}{}
\Receiveddate{}
\Accepteddate{}

 \pretitle{}


\title[Weakly interacting electrons and the renormalization group]{Weakly interacting electrons and the renormalization group}


\author[B.\ Binz]{Benedikt Binz\footnote{Corresponding
     author: E-mail: {\sf benedikt.binz@unifr.ch}, Phone: +41\,26\,300\,91\,41,
     Fax: +41\,26\,300\,97\,58}\inst{1,2}} 
\address[\inst{1}]{D\'epartement de Physique, Universit\'e de Fribourg, P\'erolles, CH-1700 Fribourg, Switzerland}
\address[\inst{2}]{Theoretische Physik, ETH Z\"urich, CH-8093 Z\"urich, Switzerland}
\author[D.\ Baeriswyl]{Dionys Baeriswyl\inst{1}}
\author[B.\ Dou\c cot]{Beno\^\i t Dou\c cot\inst{3}}
\address[\inst{3}]{Laboratoire de Physique Th\'eorique et Hautes Energies, CNRS UMR 7589, Universit\'es Paris VII, 4 Place Jussieu, 75252 Paris Cedex 05, France}

\begin{abstract}
We present a general method to study weak-coupling instabilities of a large class of interacting electron models in a controlled and unbiased way. Quite generally, the electron gas is unstable towards a superconducting state
even in the absence of phonons, since high-energy spin fluctuations create an
effective attraction between the quasi-particles. As an example, we show the occurrence of $d$-wave pairing in the repulsive Hubbard model in two dimensions. 
In one dimension or if the Fermi surface is nested, there are several 
competing instabilities. The required renormalization group formalism for this case is presented to lowest (one-loop) order on a most elementary level, connecting the idea of the ``parquet summation'' to the more modern concept of Wilson's effective action. The validity and restrictions of the one-loop approximation are discussed in detail. As a result, three different renormalization group approaches known in the literature are shown to be equivalent within the regime of applicability. We also briefly discuss the open problem of a two-dimensional Fermi system at Van Hove filling without nesting. 
\end{abstract}

\maketitle                   





\tableofcontents  

\section{Introduction}

The theoretical understanding of systems with strongly correlated electrons is one of the most important objectives of current research in physics and the key for elucidating a number of phenomena which have been observed in newly designed materials. The most spectacular example is obviously high-temperature superconductivity discovered 1986 in the cuprates, but intriguing questions have also been raised by other transition metal oxides, by organic compounds etc. 

Although the (bare) Coulomb interaction between electrons is generally strong in materials, it is important to control at least the limit of weak interactions, which is already a difficult task. Furthermore, arbitrarily weak interactions lead to strong correlations at low temperature.  It is thus worthwhile to study the weak-coupling problem thoroughly.

As will be shown in detail, na\"\i ve perturbation theory breaks down at low temperature and has to be improved by the so-called renormalization group (RG). RG concepts have been used in very different fields of modern physics and can have quite different meanings for different people. The notion was first introduced by St\"uckelberg and Petermann \cite{stuckelberg53} and independently by Gell-Mann and Low \cite{gell-mann54}  in the context of quantum field theories (such as quantum electrodynamics) in order to cope with infinities that appear in na\"\i ve perturbation theory. 

In the early 1970s, Kadanoff and Wilson have associated the RG to the procedure of mode elimination in classical statistical mechanics (or systems of bosons) \cite{wilson71,wilson74}. In Wilson's formulation, the RG transformation consists in integrating out some degrees of freedom of the system and including them in the renormalization of some parameters (for example coupling constants). This alternative formulation of the RG idea proved tremendously successful for the analysis of critical behavior in the vicinity of second order phase transitions \cite{wilson75}. 

In general, a RG transformation is some change of the length or energy scale and the RG equations describe the response of the system as the length or energy scale is changed. In the interacting electron problem considered here, the energy scale is given by the temperature or alternatively by a cutoff $\Lambda$ in the band energy of the electrons. The RG equations describe the change of a two-particle Green's function as a function of the energy scale $\Lambda$. In many cases, the RG flow to low energies produces a singularity at a finite energy scale $\Lambda_c$. This is interpreted as a transition into a strongly correlated state. The energy scale $\Lambda_c$ is then comparable to the mean-field transition temperature towards an ordered state.

It turns out that the one-loop (i.e. the leading order) approximation of the RG is equivalent to the so-called parquet approximation. This method was developed by the Soviet school \cite{pomeranchuk57} in order to treat different diverging fluctuations on an equal footing. It was successfully applied  to one-dimensional conductors \cite{Bychkov66,Dzyaloshinskii72,Gorkov74}, to the Kondo problem \cite{Abrikosov65} and to the $X$-ray absorption edge singularity problem in metals \cite{Roulet69,Nozieres69}. A detailed description of the method can be found in \cite{Roulet69}. In our derivation of the one-loop RG equations, we will follow mainly the parquet philosophy. 

The application of renormalization group (RG) ideas to fermionic problems in more than one dimension started in the mathematical physicists community \cite{Feldman90,Benfatto91} and has led to a considerable progress  during the last decade\footnote{Pedagogical introductions can be found in \cite{Polchinski92,Shankar94,froehlich96}. A precise relationship between Fermi liquid theory and the renormalization group has been established in \cite{chitov95,chitov98,Dupuis98,dupuis00}.}. For example, a  series of rigorous studies has shown that the Landau Fermi liquid theory is stable for the $D=2$ jellium model at not too low temperatures
 such that $|U\log T|<\mbox{const}$, where $U$ is the strength of the local interaction \cite{Salmhofer98,Disertori00}. For lower temperatures, the properties of the interacting system are no longer analytically connected to those of the non-interacting system. The usual interpretation is a phase transition into a superconducting state, although there exists so far no rigorous statement about the regime $T<T_c$. 

A number of different numerical schemes to compute the one-loop RG flow of $D=2$ Fermion models have been developed during the last years. They have revealed many interesting and appealing results such as the appearance of $d$-wave superconductivity in the repulsive Hubbard model close to half filling \cite{Zanchi,Zanchi00}, a spontaneous breaking of the lattice symmetry (Pomeranchuk instability) \cite{Halboth01} and a phase with suppressed uniform spin and charge susceptibilities (``insulating spin liquid'') \cite{HSFR01}. The method was further developed to a finite temperature RG scheme in \cite{HS01}. An interesting extension of the RG formalism to obtain dynamical and non-equilibrium properties of certain impurity models has been presented in Ref.~\cite{Rosch03}.

In this article, we have chosen a comparatively pedestrian approach to the weakly interacting electron problem and its RG treatment with the aim of being intuitive and didactical. The paper is organized as follows. In Section~\ref{intellat}, we show how na\"\i ve perturbation theory breaks down as a consequence of infrared divergences. In the case of a generic Fermi surface in more than one dimension, these divergences can be accounted for in a controlled way using the ladder approximation. This method is presented in Section~\ref{ladder}, where we also discuss its relation to  Kohn-Luttinger superconductivity and BCS-theory. This part is followed by a specific application in Section~\ref{superconductivity}, showing the occurrence of superconductivity in the repulsive Hubbard model in two dimensions. 

The remaining part of this article addresses the more general situation, where there is more than one competing weak coupling instability. This is generally the case in one dimension and in more than one dimension for special (nested) Fermi surfaces, which are defined in Section~\ref{nesting}.
 If the Fermi surface is nested, the ladder approximation is no longer justified and requires a non trivial extension. The necessary formalism is developed in Section~\ref{BSE} and \ref{floweq}. We first derive several exact Bethe-Salpeter equations which are then used to establish the one-loop RG equations. The region of applicability of the one-loop RG is discussed in Section~\ref{Discussion}. In Section~\ref{Wilson}, we relate our approach to the Wilsonian RG concept and we finally compare several recent RG approaches known in the literature in Section~\ref{compare}. In the Conclusion, we comment upon two-dimensional (2D) electron systems at the Van Hove filling.

Some more technical aspects are treated in four appendices. Appendix~\ref{Green} contains basic definitions and properties of the two-particle Green's function and the vertex function. In Appendix~\ref{parquetex}, we calculate a specific example of a third order parquet diagram, in order to demonstrate the general properties to which the main text refers. In Appendix~\ref{corr}, the RG formalism is extended to calculate linear response functions. Finally, some basic definitions and identities of the functional integral formulation are given in Appendix~\ref{grassmann}.

\section{Infrared divergences in na\"\i ve perturbation theory}\label{intellat}

We consider a general single-band model of interacting electrons on a $D$-dimensional periodic lattice. The Hamiltonian is of the form $H=H_0+H_I$, where
\begin{equation}
H_0=\sum_{\v k,\sigma}\epsilon_{\v k}\,c^\dagger_{\v k\sigma}c^{}_{\v k\sigma}
\label{H0}
\end{equation}
is the non-interacting part and 
\begin{equation} 
H_I=\frac1{2V}\sum_{\v k_1,\v k_2,\v k_3}g(\v k_1,\v k_2,\v k_3) \sum_{\sigma, \sigma'}c^\dagger_{\v k_1\sigma} c^\dagger_{\v k_2\sigma'}c^{}_{\v k_3\sigma'}c^{}_{\v k_1+\v k_2-\v k_3\,\sigma},\label{HI}
\end{equation}
is the most general two-body interaction respecting the global spin rotation symmetry, translational invariance and particle number conservation. $V$ is the number of lattice sites (i.e. the system volume), $c^\dagger_{\v k\sigma}$ and  $c^{}_{\v k\sigma}$ are the usual creation and annihilation operators of electrons with lattice momentum $\v k$ and spin index $\sigma=\up,\down$. The single-particle dispersion $\epsilon$ and the coupling function $g$ are smooth functions of the momenta, if the hopping matrix and the interactions are short-ranged in real space. 

Our aim is to calculate the imaginary time Green's functions. Special attention is given to the vertex function $\Gamma(k_1,k_2,k_3)$, which is defined as the one-particle irreducible part of the two-particle Green's function. The $D$+1-dimensional frequency-momentum vectors $k_i=(k_{0i},\v k_i)$, involve the Matsubara frequencies $k_{0i}$ which run over the odd multiples of $\pi/\beta$, where $\beta$ is the inverse temperature. 
$\Gamma(k_1,k_2,k_3)$ represents a scattering amplitude of two particles with incoming frequency-momentum and spin labels $(k_1,\up)$ and $(k_2,\down)$ into outgoing labels $(k_3,\down)$ and $(k_1+k_2-k_3,\up)$. A precise definition of the vertex function as well as some basic symmetry properties are given in Appendix~\ref{Green}. 

The vertex is always a regular function at high enough temperatures. At low temperatures however, the vertex can have poles. These poles are usually interpreted as the manifestation of an instability of the metallic state such as superconductivity or different kinds of density waves, since a divergence of the vertex implies a divergence of some generalized susceptibility.

Since we are concerned with weak interactions, the first approach is standard perturbation theory in $H_I$. A calculation of the vertex function to second order in perturbation theory gives
\begin{equation} 
\Gamma(k_1,k_2,k_3)=-g(\v k_1,\v k_2,\v k_3)+\mbox{PP}+\mbox{PH1}+\mbox{PH2}
\label{2ndorder}
\end{equation}
\begin{align*}
\begin{split}
\mbox{PP}&=\frac1{\beta V}\sum_p
g(\v k_1,\v k_2,\v k-\v p)C(p)C(k-p) g(\v p,\v k-\v p,\v k_3),\\
\mbox{PH1}&=\frac1{\beta V}\sum_p
g(\v k_1,\v p+\v q_1,\v k_3)C(p)C(p+q_1) g(\v p,\v k_2,\v p+\v q_1),\\
\mbox{PH2}&=\frac1{\beta V}\sum_pC(p)C(p+q_2)
\left[-2g(\v k_1,\v p,\v p+\v q_2) g(\v p+\v q_2,\v k_2,\v k_3)\right.\\ 
 &\qquad\left.+g(\v p,\v k_1,\v p+\v q_2) g(\v p+\v q_2,\v k_2,\v k_3)+
g(\v k_1,\v p,\v p+\v q_2) g(\v p+\v q_2,\v k_2,\v p)\right],
\end{split}
\end{align*}
 where  $k=k_1+k_2$, $q_1=k_3-k_1$, $q_2=k_3-k_2$ and the single-particle Green's function of the non-interacting system is given by
\begin{equation}
C(k)=\frac1{ik_0-\xi_{\v k}},
\end{equation}
 where $\xi_{\v k}=\epsilon_{\v k}-\mu$. The representation of the second order contributions in terms of Feynman diagrams is shown in Fig.~\ref{diags}. The internal electron lines stand for free electron propagators $C$ and the wavy interaction lines stand for coupling functions $g$, that depend on the incoming and outgoing momenta. The convention is that the spin index is conserved along the fermion lines. The minus sign in the first of the three PH2 diagrams comes from the fermion loop and the factor $2$ in the same diagram from the sum over the spin index in the fermion loop. PP is referred to as the particle-particle (p-p) diagram and PH1 and PH2 as particle-hole (p-h) diagrams. 

\begin{figure} 
\centerline{\includegraphics[width=11cm]{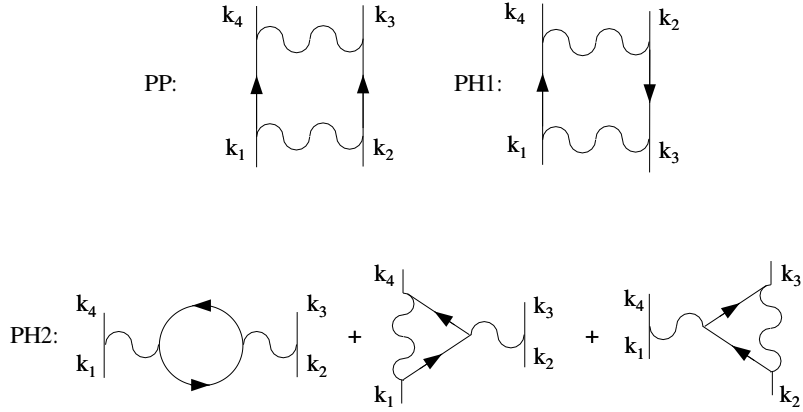}}
\caption{Second order diagrams for the one-particle irreducible vertex $\Gamma(k_1,k_2,k_3)$. $k_4$ is determined by energy-momentum conservation.}\label{diags}
\end{figure}

Suppose for a moment that $g$ is constant ($=U$ in the Hubbard model). The p-p diagram is then given by $g^2$ multiplied by the p-p bubble
\begin{eqnarray}
B^{pp}(k)&=&\frac{1}{\beta V}\sum_pC(p)C(k-p)\\
&=&\frac1{V}\sum_{\v p}\frac{n(\xi_{\v p})+n(\xi_{\v k-\v p})-1}{ik_0-\xi_{\v p}-\xi_{\v k-\v p}},
\end{eqnarray}
where $n(\xi)=(1+\exp{\beta\xi})^{-1}$ is the Fermi distribution function. This quantity is diverging at $k=0$ as the temperature goes to zero. At a finite temperature $T=\beta^{-1}$ one finds
\begin{equation}
B^{pp}(0)=\int_{-W}^W\!\ud\xi\,\nu(\xi)\,\frac{\tanh\frac{\beta\xi}2}{2\xi} =\nu(0)\,\log{\frac WT}+\mbox{``finite''},\label{logT}
\end{equation}
where $W$ is the bandwidth, $\nu(\omega)=1/V\sum_{\v k}\delta(\xi_{\v k}-\omega)$ is the density of states, and ``finite'' is any contribution with a finite limit $T\to0$. The only two assumptions are reflection symmetry $\xi_{\v k}=\xi_{-\v k}$ and that the density of states $\nu(\xi)$ is an analytic function of $\xi$ at the Fermi level. Under the same assumption it can be shown that the largest term of order $n$ in the coupling diverges as $\log^{n-1}(W/T)$. 

It is clear that na\"\i ve perturbation theory breaks down at large enough values of the coupling or low enough temperature, such that $|g\log(W/T)|\sim1$. Alternatively, one can organize the perturbation series in the form
\begin{equation} 
\Gamma=a_1\,g+a_2\,g^2+\cdots,\label{dev}
\end{equation}
where $a_1, a_2,\ldots$ are analytic functions of the product $g\log(W/T)$.
In the following we  concentrate on the leading order contributions, i.e. we calculate only the first term in this expansion. This means that diagrams of a given order in $g$ are calculated to leading logarithmic precision and all the subleading contributions such as the term ``finite'' in \eref{logT} are neglected. In other words, $g$ is considered to be small, but $g\log(W/T)$ is not.

One can regularize the zero temperature perturbation theory by introducing artificially an infrared cutoff $\Lambda$. There are several ways to do this. One method of avoiding the singularity is to remove from the theory the single-particle quantum states which are too close to the Fermi surface. This amounts to replace the bare propagator by 
\begin{equation}
C_\Lambda(k)=\Theta(|\xi_{\v k}|-\Lambda)C(k),
\end{equation}
where $\Theta$ is the Heavyside step function. In this case, the vertex function becomes cutoff-dependent and we use the notation $\Gamma_\Lambda(k_1,k_2,k_3)$. In the presence of the cutoff, the logarithms in Eqs.~\eref{logT} and \eref{dev} are  replaced by $\log{\frac W{\max\{T,\Lambda\}}}$. Alternatively, one could also use the quantity $\sqrt{k_0^2+\xi_{\v k}^2}$ instead of $|\xi_{\v k}|$ to introduce the cutoff or replace $\Theta$ by a smooth function without changing the result to leading logarithmic order.  We are thus free to calculate at zero temperature, introduce the cutoff in the most convenient way for calculations and in the end of the calculation replace $\Lambda$ by $T$. It is clear that the correspondence $\Lambda\leftrightarrow T$ is only correct to leading logarithmic order. In a more elaborate calculation including subleading contributions, the result depends on how the cutoff is introduced. 

\section{Ladder approximation and BCS theory}\label{ladder}

\begin{figure}
\centerline{\includegraphics[width=6cm]{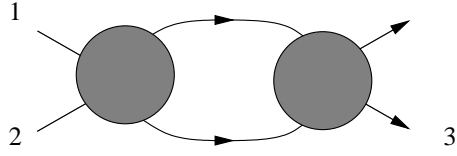}}
\caption{The structure of diagrams, which are reducible in the p-p channel. The grey circles represent any sub-diagram.}\label{ppreducible}
\end{figure}

We now calculate the vertex to leading order in the expansion \eref{dev}. In a generic situation in more than one dimension without any fine tuned parameter, the only diverging terms occur in the p-p diagram. The other second order terms have a finite limit for $T,\Lambda\to0$ except in one dimension or for special situations, which will be discussed later. 
Having identified the p-p diagram as the main source of the infrared singularity, we define the class of p-p irreducible diagrams. A diagram contributing to the two-particle scattering vertex is called reducible in the p-p channel if it is of the form shown in Fig.~\ref{ppreducible}, i.e. if it can be divided into two disconnected pieces by cutting two particle lines. In the opposite case it is called irreducible in the p-p channel or simply p-p irreducible.
 
In the following, we use the notation $I^{pp}(k_1,k_2,k_3)$ for  
the sum over all p-p irreducible diagrams. To second order in perturbation theory, this quantity is given by Eq.~\eref{2ndorder} with the singular PP contribution omitted. It can thus be calculated by na\"\i ve perturbation theory without the problem of infrared divergences.

\begin{figure}
\centerline{\includegraphics[width=7cm]{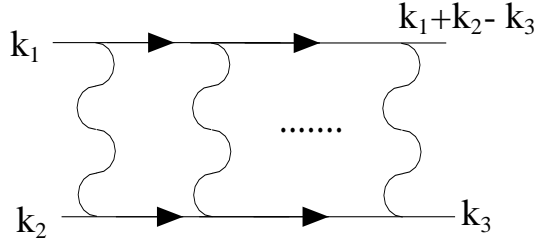}}
\caption{The structure of p-p ladder diagrams. The wavy lines stand for $-I^{pp}_\Lambda$ and the electron lines represent full propagators $G_\Lambda$, in the presence of the infrared cutoff $\Lambda$.}\label{ppladders}
\end{figure}

The exact full vertex is given in terms of the p-p irreducible part and the single-electron propagator by a summation over the so-called ladder diagrams (Fig.~\ref{ppladders}). The analytical expression of this series is 
\be
\begin{split}
\Gamma^{BCS}_{\Lambda,q}(k,k')&=I^{BCS}_{\Lambda,q}(k,k')
+\frac1{\beta V}\sum_pI^{BCS}_{\Lambda,q}(k,p)D^{pp}_{\Lambda,q}(p)\,I^{BCS}_{\Lambda,q}(p,k')\\
&\quad+\frac1{(\beta V)^2}\sum_{p_1,p_2}I^{BCS}_{\Lambda,q}(k,p_1)D^{pp}_{\Lambda,k}(p_1)\,I^{BCS}_{\Lambda,q}(p_1,p_2)D^{pp}_{\Lambda,k}(p_2)\,I^{BCS}_{\Lambda,q}(p_2,k')+\cdots,
\end{split}\label{seriespp}
\end{equation}
where, $q$ is the total frequency-momentum of the incoming and outgoing pair of particles, $D^{pp}_{\Lambda,q}(p)=G_\Lambda(p)G_\Lambda(q-p)$ is the propagator of the pair and we have defined
\begin{eqnarray}
\Gamma^{BCS}_{\Lambda,q}(k,k')&=&\Gamma_{\Lambda}(k,q-k,q-k'),\nonumber\\
I^{BCS}_{\Lambda,q}(k,k')&=&I^{pp}_{\Lambda}(k,q-k,q-k').
\end{eqnarray}
In principle, the internal electron propagators in Fig.~\ref{ppladders} are
exact single-particle Green's functions  $G_\Lambda(p)=\Theta(|\xi_{\v
  p}|-\Lambda)/(ip_0-\xi_{\v p}-\Sigma_\Lambda(p))$, where
$\Sigma_\Lambda(p)$ is the self-energy in the presence of the
cutoff. Self-energy corrections have three main effects. First they change the
shape and location of the Fermi surface, second they modify the properties of
the single particle dispersion (the Fermi velocity) and third they lead to a
reduction of the quasi-particle weight. The deformation of the Fermi surface
requires in general the introduction of counter-terms\footnote{See Section~5.7
  of Nozi\`eres' book \cite{Nozieres} for a comprehensive
  explanation. Rigorous mathematical statements about the moving Fermi surface
  have been presented in \cite{Feldman96,Feldman98,Feldman99}. For recent
  investigations of interaction-induced Fermi surface deformations, see 
\cite{dusuel03,ledowski03,neumayr03,ferraz03}.}. We will ignore this effect and assume that in the weak coupling limit, the interacting Fermi surface is not so different from the non interacting one such that the difference is not crucial. Perturbative corrections to the Fermi velocity $\nabla_{\v p}\Sigma(p)$ and the quasi-particle weight $z=(1+i\partial_{p_0}\Sigma(p))^{-1}$ are finite as $T\to0$ and therefore do not contribute to leading logarithmic order. We will therefore neglect self-energy corrections in the remaining part of this section. 

As argued above, the p-p irreducible vertex function $I^{BCS}_{\Lambda,q}$ can be calculated within na\"\i ve perturbation theory. Let us thus focus on how to sum the infinite series Eq.~\eref{seriespp}, for given $I^{BCS}_{\Lambda,q}$. 

For simplicity, we set the frequency $q_0$ equal to zero and neglect the frequency-dependence of $I^{BCS}(k,k')$. This point will be justified later. We focus on small values of $\v q$ and $\Lambda$, because this is where the infrared divergence appears. The p-p irreducible vertex has no infrared singularity  and can thus safely be replaced by 
\be
I^{BCS}(\v k,\v k')=\lim_{\Lambda,\v q\to0}I^{BCS}_{\Lambda,\v q}(\v k,\v k').
\end{equation} 
In this way, $I^{BCS}$ becomes a real-valued and symmetric function of the momenta $\v k$ and $\v k'$. One can then define a Hermitian operator\footnote{To be more precise, the operator defined in Eq.~\eref{operator} is Hermitian with respect to the scalar product given in Eq.~\eref{orthogonality}.}, which acts on a function $f(\v k)$ as
\be
f(\v k)\longrightarrow\frac1{VB^{pp}_\Lambda(\v q)}\sum_{\v k'}I^{BCS}(\v k,\v k')\tilde D^{pp}_{\Lambda,\v q}(\v k')f(\v k'),\label{operator}
\end{equation}
where $\tilde D^{pp}_{\Lambda,\v q}(\v k)=1/\beta\sum_{k_0}D^{pp}_{\Lambda,\v q}(k)$ is the non-interacting frequency-summed pair propagator and
$B^{pp}_\Lambda(\v q)$ 
is the p-p bubble in the presence of the cutoff. The factor $1/B^{pp}_\Lambda(q)$ has been introduced  in such a way, that Eq.~\eref{operator} acquires a finite limit for $\Lambda,\v q\to0$.

This implies that $I^{BCS}$ has a spectral representation given by
\be
I^{BCS}(\v k,\v k')=\sum_n\lambda_{n}\ e_{n}(\v k)e_{n}(\v k'),\label{spectral}
\end{equation}
where $\lambda_n$ are the eigenvalues of the operator \eref{operator} in the limit $\Lambda,\v q\to0$ and $e_{n}(\v k)$ are the corresponding eigenfunctions. The latter satisfy the orthogonality relation
\be
\frac1{VB^{pp}_\Lambda(\v q)}\sum_{\v k}e_{n}(\v k)\tilde D^{pp}_{\Lambda,\v q}(\v k)e_{m}(\v k)=\delta_{nm}\label{orthogonality}
\end{equation}
in the limit of small $\v q$ and $\Lambda$. Using Eqs.~\eref{spectral} and \eref{orthogonality}, Eq.~\eref{seriespp} reduces to a simple geometric series and is easily summed, giving 
\be
\Gamma^{BCS}_{\Lambda,\v q}(\v k,\v k')=\sum_n\frac{\lambda_{n}}{1-\lambda_{n}B^{pp}_\Lambda(\v q)}\ e_{n}(\v k)e_{n}(\v k').\label{GBCS}
\end{equation}
Clearly,  positive eigenvalues produce a singularity in the vertex when $\Lambda$ is lowered to the energy scale $\Lambda_c$ given by $\lambda_nB^{pp}_{\Lambda_c}(0)=1$, or equivalently
\be
\Lambda_c=W\exp\left(-\frac1{\nu(0)\lambda_n}\right).\label{Lc}
\end{equation}

The divergence of the vertex $\Gamma^{BCS}_\Lambda$ at $\Lambda=\Lambda_c$ is usually interpreted as signaling the onset of superconductivity. In fact, the divergence is caused by low energy p-p pairs and requires some attractive channel in the interaction (corresponding to a positive eigenvalue of $I^{BCS}$). This suggests that the instability is associated with the formation of Cooper pairs. Moreover formula  \eref{Lc} for the critical energy scale corresponds exactly to the BCS formula for the critical temperature. To establish the connection to BCS-theory, consider the standard gap equation
\be
\Delta_{\v k}=-\frac1{V}\sum_{\v k'}g^{BCS}(\v k,\v k')\frac{\Delta_{\v k'}}{2E_{\v k'}}\tanh(\frac{\beta E_{\v k'}}2),
\end{equation}
where $g^{BCS}(\v k,\v k')$ is the coupling function, which describes the scattering of Cooper pairs from momenta $\v k$ and $-\v k$ towards $\v k'$ and $-\v k'$. Close to the critical temperature, the mean-field dispersion $E_{\v k}=\sqrt{\Delta_{\v k}^2+\xi_{\v k}^2}$ is replaced by $|\xi_{\v k}|$. The resulting equation coincides exactly with the eigenvalue equation of the operator \eref{operator}
 at $\v q=\v 0$ and $\Lambda=\Lambda_c$, if we identify $-g^{BCS}(\v k,\v k')$ with $I^{BCS}(\v k,\v k')$ and $\Delta_{\v k}$, up to a normalizing factor, with $e_n(\v k)$. We conclude that the eigenfunction $e_n(\v k)$ with the largest positive eigenvalue determines the momentum-dependence ($s$-wave, $p$-wave, etc.) of the superconducting gap.

It is actually not necessary to calculate the complete function $I^{BCS}(\v k,\v k')$, but only its value on the Fermi surface. In fact, the eigenvalue equation reduces at small energy scales to
\be
\lambda_n\,e_n(\v k)=\frac1{V \nu(0)}\sum_{\v k'}\delta(\xi_{\v k'})\,I^{BCS}(\v k,\v k')\,e_n(\v k') + O(\frac1{\log \Lambda}).\label{eveq}
\end{equation}
Thus, the eigenvalues $\lambda_n$ and $\left.e_n(\v k)\right|_{\xi_{\v k}=0}$ are determined by the value of $I^{BCS}(\v k,\v k')$ on the Fermi surface only. In this sense, the dependence of $I^{BCS}$ on the energies $\xi_{\v k}$ and $\xi_{\v k'}$ can be classified as irrelevant. The same is true for the dependence on the frequencies $k_0$ and $k'_0$, which has been neglected. Only the dependence on the momenta moving along the Fermi surface is crucial.

\section{Superconductivity from repulsive interactions}\label{superconductivity}

As an example, we will now calculate the superconducting instabilities for an extended 2D Hubbard model away from half-filling\footnote{Similar calculations were performed before by Zanchi and Schulz \cite{Zanchi96} and more recently by Guinea et al.~\cite{guinea02}. Our results are in perfect agreement with theirs.}. The electron hopping is restricted to nearest neighbors, i.e. $\xi_{\v k}=-2t(\cos k_x+\cos k_y)-\mu$. Fig.~\ref{genFS} shows the Fermi surface for a typical electron density. The variable $\theta$ used to parameterize the Fermi surface is defined as the radial angle. 
The interaction
\be
H_I=U\sum_{\v r} n_{\up \v r} n_{\down \v r}+V\sum_{\langle\v 
r,\v r'\rangle}n_{\v r}n_{\v r'}
\end{equation}
 contains a nearest neighbor term  $V$ in addition to the Hubbard $U$. This corresponds to a coupling function $g(\v k_1,\v k_2,\v k_3)=U+2V[\cos(k_{3,x}-k_{2,x})+\cos(k_{3,y}-k_{2,y})]$.

\begin{figure}  
 \centerline{\includegraphics[width=4cm]{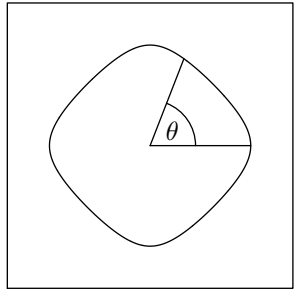}}\vspace{-11pt}
\centerline{\hspace{0.6cm}\raisebox{2.1cm}[0pt][0pt]{$\theta$}}
\caption{The Fermi surface of the nearest-neighbor tight-binding band for $\mu=-0.8$ (i.e. $n\approx0.7$).}\label{genFS}
\end{figure}

We study the superconducting instabilities in two steps. First we calculate within perturbation theory the p-p irreducible vertex $I(\theta,\theta')=I^{BCS}_{\v q=\v 0,\Lambda=0}(\v k(\theta),\v k(\theta'))$, where $\v k(\theta)$ is the Fermi momentum corresponding to the angle $\theta$. A  discretization of the $\theta$ variable into 32 patches is introduced to allow for a numerical treatment. The second step is the solution of the eigenvalue equation \eref{eveq}. This is a trivial numerical operation, once the $\theta$-variable has been discretized. 

 The eigenfunctions $e_n(\theta)$ can be characterized by their transformation properties with respect to the symmetry group of the square lattice. The $D_4$ point group has five irreducible representations: $A_1$,~$A_2$,~$B_1$, $B_2$ and $E$. They are illustrated in Table \ref{irreps}.

\begin{table}[ht]
\centerline{\begin{tabular}{c|l|l}
Irreducible representation&Basis&Simple cases \\
\hline
$A_1$&$\cos{4m\theta}$& $s$-wave \\
$A_2$&$\sin{4m\theta}$& - \\
$B_1$&$\cos{(4m+2)\theta}$&$d_{x^2-y^2}$\\
$B_2$&$\sin{(4m+2)\theta}$&$d_{xy}$\\
$E$&$\left\{\begin{array}{c}\cos{(2m+1)\theta}\\ \sin{(2m+1)\theta}\end{array}\right.$&$p$-wave, $f$-wave
\end{tabular}}
\caption{Irreducible representations of the $D_4$ point group, a basis of functions, which span the corresponding subspace ($m$ is an arbitrary integer) and the term commonly assigned to the simplest order parameters in each case. The $A_2$ subspace consists only of rapidly oscillating functions with at least eight nodes along the Fermi surface.}\label{irreps}
\end{table}

All the calculations have been done for a repulsive on-site interaction $U=t$. In the case of an attractive $V<0$, the appearance of superconductivity is not surprising. In fact, the first order vertex $I=-g$ has positive eigenvalues in this case. The results can be seen in Fig.~\ref{Vneg}. They show triplet $p$-wave superconductivity for electron densities $n<0.65$ and singlet  $d_{x^2-y^2}$-wave superconductivity for $0.65<n$ (close to half filling, additional instabilities arise in the p-h channel).  The second order contributions to $I(\theta,\theta')$ give only minor changes to the present result, although they introduce small positive eigenvalues in the symmetry blocks $B_2, A_1$ and $A_2$.

\begin{figure}   
\centerline{\includegraphics[width=9cm]{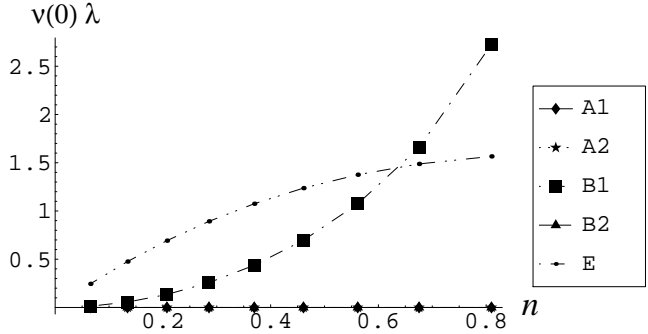}}
\caption{The positive eigenvalues of $I(\theta,\theta')$ multiplied by the density of states at the Fermi level as a function of the electron density for an attractive nearest-neighbor interaction $V=-0.1t$ and $U=t$, as obtained by a first order calculation. The only positive eigenvalues are in the $E$ and $B_1$ symmetry block. The corresponding eigenfunctions are $p$-wave like (for $E$) and $d_{x^2-y^2}$-wave like (for $B_1$). The quantity $\nu(0)\lambda$ determines the energy scale of the superconducting instability (see Eq.~\eref{Lc}).}\label{Vneg}
\end{figure}

In contrast, in the pure Hubbard model ($V=0$, $U>0$), positive eigenvalues of $I(\theta,\theta')$ appear only via second order terms. 
Since the sum of the three PH2 diagrams is zero in the Hubbard model, the only second order contribution to $I(\theta,\theta')$ comes from the PH1 diagram. The latter is computed numerically, using a mesh of $3000\times3000$ $\v k$-points in the Brillouin zone and a temperature $T=0.001 t$. Kohn and Luttinger have investigated the same diagram in their historical paper \cite{Kohn65}. The difference is that we are dealing with an anisotropic 2D system as opposed to the isotropic 3D case considered in \cite{Kohn65}.

 The results in Fig.~\ref{Vzero} for the Hubbard model show $d_{xy}$-wave symmetry for $n<0.5$ and $d_{x^2-y^2}$-wave for  $n>0.6$. In between there is a narrow range of densities where triplet superconductivity (with an $f$-wave like order parameter) is favored. Note that the eigenvalues of Fig.~\ref{Vzero} created by second order terms are in general smaller than those in Fig \ref{Vneg} with an attractive $V$, leading to superconductivity at lower energy scales.

\begin{figure}   
\centerline{\includegraphics[width=13cm]{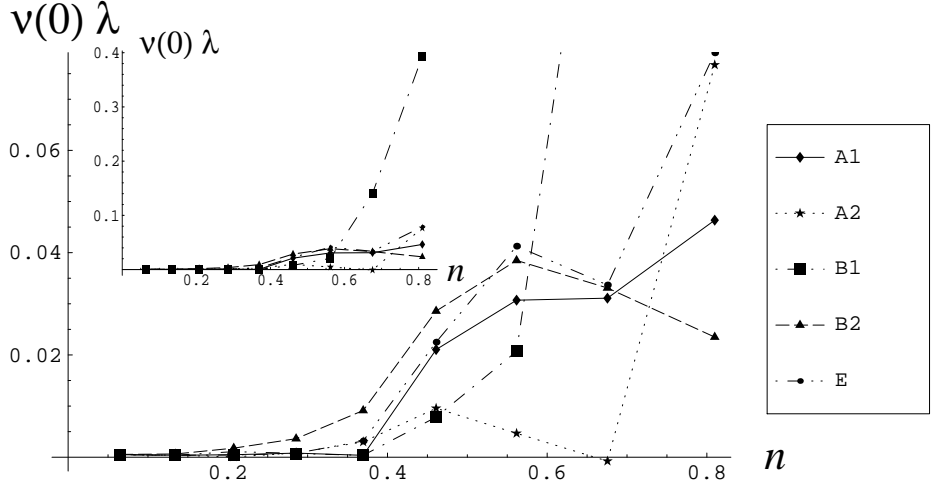}}
\caption{Same as Fig.~\ref{Vneg} from a second order calculation for the pure Hubbard model $V=0$ and $U=0.5t$ (Kohn-Luttinger superconductivity). The largest eigenvalue of each symmetry block is shown. The corresponding eigenfunctions are $d_{xy}$-wave like for $B_2$ ($n<0.5$) and $d_{x^2-y^2}$-wave like for $B_1$ ($n>0.6$), and approximately $f$-wave for $E$ (which is leading only in a narrow range of dopings). The inset shows the global behavior with its drastic increase of the $d_{x^2-y^2}$-wave eigenvalue close to half-filling.}\label{Vzero}
\end{figure} 

The appearance of superconductivity by second-order corrections to the vertex has a physical interpretation in terms of an effective attractive interaction, which is mediated by spin fluctuations. In fact, $I(\theta,\theta')$ equals
\be
I(\theta,\theta')=-U-U^2\chi_0(\v k_{\theta}+\v k_{\theta'}),
\end{equation}
where $\chi_0(\v q)$ is the spin susceptibility of the non-interacting system. The attractive part of $I(\theta,\theta')$ comes therefore from the spin susceptibility $\chi_0(\v q)$. It is thus fair to say that an effective attraction is created by the exchange of spin fluctuations. The situation is analogous to the conventional superconductivity of metals, where an effective attraction is created through the exchange of phonons, however with the important difference that the same electrons which feel the effective interaction are also responsible for the spin fluctuations.  

The idea of spin-fluctuation-induced pairing has led to semi-phenomenological theories of superconducting materials with strong magnetic correlations such as the cuprates which are antiferromagnets in the absence of doping \cite{BickersPRL89,Bickers89,Moriya90,Monthoux92,Kondo99}.

These results strongly support that there is superconductivity in the repulsive Hubbard model at weak coupling. While for electron densities smaller than $0.6$ we expect superconductivity only at exponentially small temperatures, the energy scale for $d_{x^2-y^2}$- pairing increases drastically as the system approaches half-filling.  This effect is partly due to the enhanced density of states, but   approximate nesting, which enhances spin fluctuations and thus the effective attraction among electrons, is likely to be more important. Very close to half filling, spin fluctuations become so strong that the ladder approximation is no longer valid, and a consistent treatment of both p-p and p-h diagrams is required. This is the subject of the remaining part of this paper.

\section{Nesting}\label{nesting}

In Section~\ref{ladder} it was assumed that the divergences in the perturbation series of the vertex are exclusively generated by the p-p diagram. 
The divergence in the p-p diagram at $\v k=0$ depends on the parity relation $\xi_{\v p}=\xi_{-\v p}$, which is applicable quite generally. 

In contrast, divergences in the p-h diagrams arise only if the Fermi surface has the so-called nesting property. A Fermi surface $FS$ is called p-h nested if the following conditions are satisfied.
\begin{enumerate} 
\item There exists a vector $\v Q$ such that  $FS$ has some finite overlap\footnote{By ``finite overlap'', we mean an overlap with a finite $(D-1)$-dimensional volume. Thus a single overlapping point is sufficient for $D=1$ (and ladder systems), for $D=2$ it should be a curve of finite length, etc.} with its own translation $\v Q+FS$.
\item The overlap is such that the occupied part ($\xi_{\v p}<0$) in the vicinity of $FS$ is covering the empty region ($\xi_{\v p}>0$) of $\v Q+FS$ and vice versa. 
\end{enumerate} 
In this case the p-h diagram PH1 shows a logarithmic divergence for $\v q_1=\v Q$ and PH2 is divergent for $\v q_2=\v Q$. For example every bipartite lattice (i.e. a lattice with electrons hopping only from one sub-lattice $A$ to another sub-lattice $B$) leads to a p-h nested Fermi surface at half filling.

\bigskip Similarly one can define p-p nesting as follows.
\begin{enumerate} 
\item There exists a vector $\v K$ such that  $FS$ has some finite overlap with $\v K-FS$.
\item  The overlap is such that the occupied (empty) side of $FS$ is covering the occupied (empty) side of $\v K-FS$.
\end{enumerate} 
In this case, the p-p diagram PP shows a logarithmic divergence for $\v k=\v K$.  For example, every Fermi surface with spatial inversion symmetry is p-p nested for $\v K=\v 0$.

These are necessary conditions for the occurrence of infrared divergences if the density of states at the Fermi level is finite, i.e. in the absence of Van Hove singularities.

The situation is particularly rich if the Fermi surface has flat portions. In this case there is usually a continuum of nesting vectors satisfying the above conditions. RG or parquet calculations for this case are provided in Refs.~\cite{Zheleznyak97,Abreu01,Dusuel01}.

\section{Bethe-Salpeter equations}\label{BSE}

If the divergences arise in both  p-p and p-h bubbles, the ladder approximation used in Section~\ref{ladder} is not sufficient.  In fact, in such cases the leading term of the perturbative expansion \eref{dev} is given by the so-called parquet diagrams. These diagrams are obtained by retaining the five one-loop (or second order) diagrams shown in Fig.~\ref{diags} and by replacing any bare vertex by one of the one-loop diagrams and continuing this process to any order. An example is given in Appendix~\ref{parquetex}.

To keep track of all the logarithmic divergences, one defines three different ``channels'' of two-particle reducible diagrams. A diagram is called reducible in the p-p channel if it has the structure shown in Fig.~\ref{ppreducible}. Similarly, the diagrams of the form shown in Fig.~\ref{reducible} a and b are called reducible in the channels p-h 1 and p-h 2, respectively.

\begin{figure}  
\centerline{\includegraphics[width=6cm]{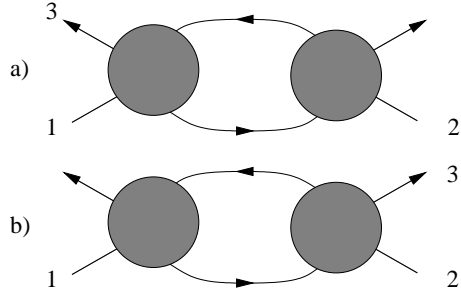}}
\caption{The structure of diagrams, which are reducible in  the p-h 1 channel (a) and the p-h 2 channel (b). The grey circles represent any sub-diagram.}\label{reducible}
\end{figure}

Let us denote by

\begin{tabular}{ll}
$R^{pp}_\Lambda(k_1,k_2,k_3)$&the set of reducible diagrams in the p-p channel, with an infrared cutoff $\Lambda$,\\
$R^{ph1}_\Lambda(k_1,k_2,k_3)$&the set of reducible diagrams in the p-h 1 channel,\\
$R^{ph2}_\Lambda(k_1,k_2,k_3)$&the set of reducible diagrams in the p-h 2 channel,\\
$I_\Lambda(k_1,k_2,k_3)$&the set of two-particle irreducible diagrams.
\end{tabular}\\

It is rather simple to check that a given diagram is either two-particle irreducible or reducible in only one of the three possible channels p-p, p-h 1 and p-h 2. Hence
\begin{equation}
\Gamma_\Lambda=I_\Lambda+R^{pp}_\Lambda+R^{ph1}_\Lambda+R^{ph2}_\Lambda.\label{decomposition}
\end{equation}
The set of irreducible graphs in each of the three channels $\diamond=pp, ph1$ or $ph2$ is defined by  $I^\diamond_\Lambda=\Gamma_\Lambda-R^\diamond_\Lambda$. We already encountered $I^{pp}$ in Section~\ref{ladder}, where it could be calculated within perturbation theory. This is no longer true in the present case, where infrared divergences occur in the p-h channels as well.

Clearly, $R^{pp}_\Lambda$ is given by the ladder series Eq.~\eref{seriespp}, but without the first (zero bubble) term. 
 The first-order term is missing since it corresponds to an irreducible contribution. We write this series formally as
\begin{equation}
R^{pp}_\Lambda=I^{pp}_\Lambda D^{pp}_\Lambda I^{pp}_\Lambda+I^{pp}_\Lambda D^{pp}_\Lambda I^{pp}_\Lambda D^{pp}_\Lambda I^{pp}_\Lambda+\ldots\ .\label{Rppseries}
\end{equation}
It satisfies the integral equation
\begin{equation}
R^{pp}_\Lambda=I^{pp}_\Lambda D^{pp}_\Lambda I^{pp}_\Lambda+I^{pp}_\Lambda D^{pp}_\Lambda R^{pp}_\Lambda
\end{equation}
or, using the relation $\Gamma=I^{pp}+R^{pp}$,
\begin{equation}
R^{pp}_\Lambda=I^{pp}_\Lambda D^{pp}_\Lambda\, \Gamma_\Lambda.\label{formalpp}
\end{equation}
This is the Bethe-Salpeter equation\footnote{See for example \cite{Abrikosov,Nozieres}. The  Bethe-Salpeter equation was originally introduced to calculate two-particle bound-states within quantum electrodynamics \cite{Salpeter51}.}. With all the functional dependences included it reads
\begin{equation}
R^{pp}_\Lambda(k_1,k_2,k_3)=\frac1{\beta V}\sum_pI^{pp}_\Lambda(k_1,k_2,k-p)D^{pp}_{\Lambda,k}(p)\,\Gamma_\Lambda(p,k-p,k_3),\label{BSpp}
\end{equation}
where $k=k_1+k_2$ and $D^{pp}_{\Lambda,k}(p)=G_\Lambda(p)G_\Lambda(k-p)$.

\begin{figure}  
\centerline{\includegraphics[width=7cm]{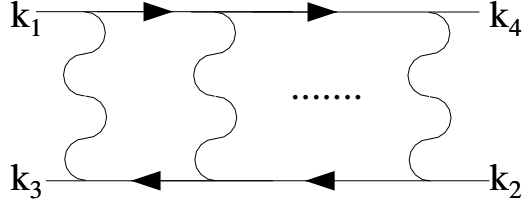}}
\caption{The p-h ladder diagrams. The wavy lines stand for $-I^{ph1}_\Lambda$ and the electron lines represent full propagators $G_\Lambda$.}\label{ph1ladders}
\end{figure}
\begin{figure}  
\centerline{\includegraphics[width=12cm]{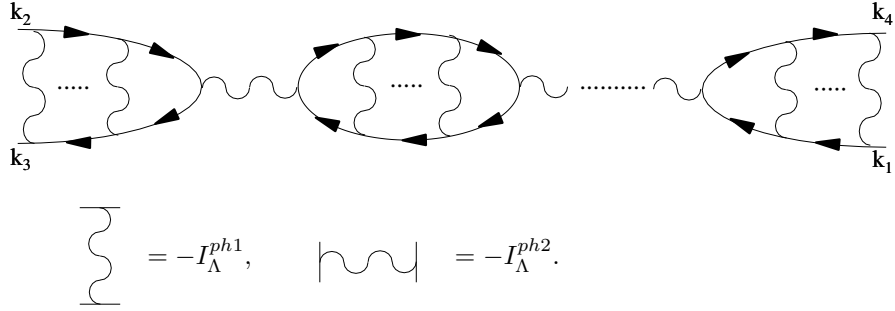}}\vspace{-11pt}
\centerline{\hspace{-2.7cm}\raisebox{1.2cm}[0pt][0pt]{$=-I_\Lambda^{ph1},$}\hspace{2.7cm}\raisebox{1.2cm}[0pt][0pt]{$=-I_\Lambda^{ph2}.$}}
\caption{The general structure of a reducible diagram in the p-h 2 channel. The wavy lines drawn horizontally stand for $-I^{ph2}_\Lambda$, but the wavy lines drawn vertically stand for $-I^{ph1}_\Lambda$. The electron lines stand for full propagators $G_\Lambda$.}\label{ph2ladders}
\end{figure}

There are similar equations for the two other channels. First, $R^{ph1}$ is given by a series over the p-h ladders shown in Fig.~\ref{ph1ladders}, where the wavy lines stand for $-I^{ph1}_\Lambda$ and the electron lines denote the full propagators $G_\Lambda$. In complete analogy with the p-p case one finds the Bethe-Salpeter equation for the p-h 1 channel
\begin{equation}
R^{ph1}_\Lambda(k_1,k_2,k_3)=\frac1{\beta V}\sum_pI^{ph1}_\Lambda(k_1,p+q_1,k_3)D^{ph}_{\Lambda,q_1}(p)\,\Gamma_\Lambda(p,k_2,p+q_1),\label{ph1BS}
\end{equation}
where $q_1=k_3-k_1$ and 
\begin{equation}
D^{ph}_{\Lambda,q}(p)=G_\Lambda(p)G_\Lambda(p+q).
\end{equation}
The p-h 2 channel is more involved. In fact, a general $R^{ph2}$-diagram has the structure shown in Fig.~\ref{ph2ladders}, where the wavy lines drawn horizontally stand for $-I^{ph2}_\Lambda$, but the wavy lines drawn vertically correspond to $-I^{ph1}_\Lambda$. The electron lines stand for full propagators $G_\Lambda$. Note that the diagram must have at least one horizontal wavy line (in the opposite case it is a p-h ladder and not reducible in the p-h 2 channel).

The simplest examples of p-h 2-reducible diagrams are shown in Fig.~\ref{diags}. The first PH2 diagram of this figure is given by 
\begin{equation}
-2\ \frac1{\beta V}\sum_pI^{ph2}_\Lambda(k_1,p,p+q_2)D^{ph}_{\Lambda,q_2}(p)\,I^{ph2}_\Lambda(p+q_2,k_2,k_3),\label{explicitph2}
\end{equation}
where $q_2=k_3-k_2$. We will write this specific convolution of the functions  $I^{ph2}_\Lambda$ and  $D_\Lambda^{ph}$ as a formal product 
\begin{equation}
-2\ I^{ph2}_\Lambda D_\Lambda^{ph} I^{ph2}_\Lambda. \label{formalph2}
\end{equation}
Note that this notation differs from the formal product introduced before in the p-p case. With this notation one can write
\begin{equation}
R^{ph2}= -2\ I^{ph2} D^{ph} I^{ph2}+(XI^{ph1}) D^{ph} I^{ph2}+I^{ph2} D^{ph} (XI^{ph1})+\mbox{``higher order terms'',}\label{2ndorderph2}
\end{equation}
where the overall $\Lambda$-index has been omitted for the simplicity of notation and $XF(k_1,k_2,k_3)=F(k_2,k_1,k_3)$ for any function of three energy-momenta. The p-h ladder series can also be written using the same notation:
\begin{equation}
XR^{ph1}=(XI^{ph1}) D^{ph} (XI^{ph1}) +\mbox{``higher order terms''.}\label{2ndorderph1}
\end{equation}
The following definitions turn out to be very useful. 
\begin{eqnarray}
R^c&=&2R^{ph2}-XR^{ph1},\nonumber\\
I^c&=&2I^{ph2}-XI^{ph1},\label{cdef}\\
\Gamma^c&=&I^c+R^c=(2-X)\Gamma,\nonumber
\end{eqnarray}
and
\begin{eqnarray}
R^s&=&-XR^{ph1},\nonumber\\
I^s&=&-XI^{ph1},\label{sdef}\\
\Gamma^s&=&I^s+R^s=-X\Gamma.\nonumber
\end{eqnarray}
The superscripts $c$ and $s$ refer to charge and spin, respectively. Note that $\Gamma^{c/s}=\Gamma^{\up\up}\pm\Gamma^{\up\down}$, so these vertex functions enter naturally in the calculation of charge- and spin- response functions. 

Equations \eref{2ndorderph2} and  \eref{2ndorderph1} can now be written in the simple form 
\begin{equation}
R^{s,c}=-I^{s,c}D^{ph}I^{s,c}+\mbox{``higher order terms''.}
\end{equation}
The exact expressions for $R^{s,c}$ are given by the infinite series
\begin{equation}
R^{s,c}=-I^{s,c}D^{ph}I^{s,c}+I^{s,c}D^{ph}I^{s,c}D^{ph}I^{s,c}-\cdots\ .
\label{scseries}
\end{equation}
In order to prove Eq.~\eref{scseries},  we first write $R^{ph2}$ as an infinite series $R^{ph2}=\sum_{n=1}^\infty R^{ph2}_n$, where $R^{ph2}_n$ is the set of diagrams, which are exactly $n$-times reducible in the p-h 2 channel. A similar decomposition is done for $R^{ph1}$. The recursion relations between $R^{ph1}_{n+1},R^{ph2}_{n+1}$ and  $R^{ph1}_{n},R^{ph2}_{n}$ are shown graphically in Figs.~\ref{ph2recursion} and \ref{ph1recursion}. The relation for $R^{ph2}_{n}$ (Fig.~\ref{ph2recursion}) can be understood as follows. A general $R^{ph2}_{n+1}$-diagram as shown in Fig.~\ref{ph2ladders} can either start on the left with a vertical or a horizontal wavy line. If it starts with a vertical line, then it is given by the second diagram of Fig.~\ref{ph2recursion}. If it starts with a horizontal line, there are two cases. If there are no other horizontal lines except for the starting one, the diagram is given by the third term of Fig.~\ref{ph2recursion}. If there is more than one horizontal line in the diagram, it is given by the first diagram in Fig.~\ref{ph2recursion}. Analytically the equations depicted in Figs.~\ref{ph2recursion} and \ref{ph1recursion} read
\begin{eqnarray} 
R_{n+1}^{ph2}&=& -2\ I^{ph2} D^{ph} R_n^{ph2}+XI^{ph1} D^{ph} R_n^{ph2}+I^{ph2} D^{ph} XR_n^{ph1}\\
XR_{n+1}^{ph1}&=&XI^{ph1} D^{ph} R_n^{ph1},
\end{eqnarray}
or, with the definitions \eref{cdef} and \eref{sdef},
\begin{equation}
 R_{n+1}^{s,c}=-I^{s,c} D^{ph} R_n^{s,c}.
\end{equation}
From this, it is easy to deduce Eq.~\eref{scseries}. In analogy to the p-p case, we find now the Bethe-Salpeter equations for $\Gamma^s$ and $\Gamma^c$ 
\begin{equation}
R^{s,c}=-I^{s,c}D^{ph}\Gamma^{s,c}.\label{scBS}
\end{equation}

\begin{figure}  
\centerline{\includegraphics[width=12cm]{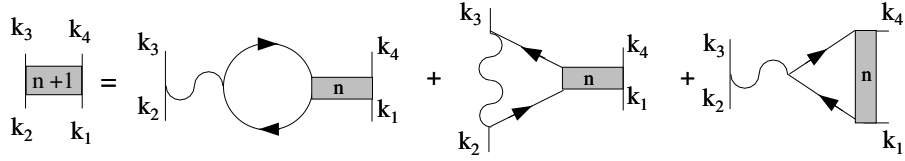}}
\caption{The recursion relation for $R^{ph2}_{n+1}$. Wavy lines drawn horizontally stand for $I^{ph2}$ and wavy lines drawn vertically stand for $I^{ph1}$. The grey rectangles with index $n$ stand for $R^{ph2}_n$ ($R^{ph1}_n$) if they are drawn horizontally (vertically).}\label{ph2recursion}
\end{figure}

\begin{figure}  
\centerline{\includegraphics[width=6cm]{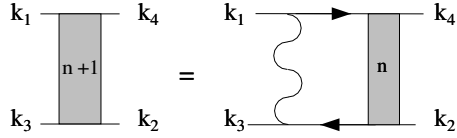}}
\caption{The recursion relation for $R^{ph1}_{n+1}$. The conventions are as in Fig.~\ref{ph2recursion}}\label{ph1recursion}
\end{figure}

\section{RG flow of the vertex}\label{floweq}

The Bethe-Salpeter equations \eref{BSpp} and \eref{scBS}, three integral equations for the three unknown functions $R^{pp}$, $R^{c}$ and $R^{s}$, are in general difficult to solve. We will address the more modest task of calculating the leading term in the perturbative expansion \eref{dev}. This will be accomplished via the flow equation, i.e. a differential equation for $\partial_\Lambda\Gamma$ where we keep only the leading terms in $\Lambda\to0$. Within this approximation, the two-particle irreducible vertex $I$ is given by the bare interaction $-g$. Thus by Eq.~\eref{decomposition},
\begin{equation}
\dot\Gamma(k_1,k_2,k_3)=\dot R^{pp}(k_1,k_2,k_3)+\dot R^{ph1}(k_1,k_2,k_3)+\dot R^{ph2}(k_1,k_2,k_3),\label{master}
\end{equation}
where the dot means a partial derivative with respect to $\Lambda$. 

Consider the first term $\dot R^{pp}$. The derivative of the Bethe-Salpeter equation \eref{formalpp} has three contributions
\begin{equation}
 \dot R^{pp}=\dot I^{pp}D^{pp}\Gamma+I^{pp}\dot D^{pp}\Gamma+I^{pp}D^{pp}\dot \Gamma.\label{Rppderiv}
\end{equation} 
Although $\dot I^{pp}$ is by no means negligible by itself (since it contains terms which are reducible in the p-h channels), the first contribution, $\dot I^{pp}D^{pp}\Gamma$, can be shown to be of subleading order in $\Lambda$. We will not prove this here, but an example is discussed in detail in Appendix~\ref{parquetex}. The last term is written as $I^{pp}D^{pp}\dot I^{pp}+I^{pp}D^{pp}\dot R^{pp}$ and $I^{pp}D^{pp}\dot I^{pp}$ is neglected for the same reason as $\dot I^{pp}D^{pp}\Gamma$. Therefore
\begin{equation}
 \dot R^{pp}=I^{pp}\dot D^{pp}\Gamma+I^{pp}D^{pp}\dot R^{pp}.
\end{equation} 
This equation can be iterated to give
\begin{eqnarray}
 \dot R^{pp}&=&I^{pp}\dot D^{pp}\Gamma+I^{pp}D^{pp}I^{pp}\dot D^{pp}\Gamma+I^{pp}D^{pp}I^{pp}D^{pp}I^{pp}\dot D^{pp}\Gamma+\cdots\nonumber\\
&=&I^{pp}\dot D^{pp}\Gamma+R^{pp}\dot D^{pp}\Gamma\nonumber\\
&=&\Gamma\dot D^{pp}\Gamma\label{rg1},
\end{eqnarray}
or written out with the full functional dependencies
\begin{equation}
\dot R^{pp}_\Lambda(k_1,k_2,k_3)=\frac1{\beta V}\sum_p\Gamma_\Lambda(k_1,k_2,k-p)\dot D^{pp}_{\Lambda,k}(p)\,\Gamma_\Lambda(p,k-p,k_3),\label{RGpp}
\end{equation}
where $k=k_1+k_2$ is the total frequency-momentum.

\begin{figure}
\centerline{\includegraphics[width=7cm]{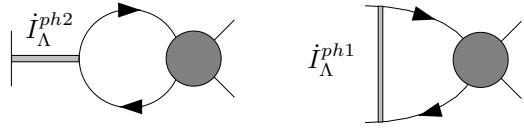}}\vspace{-11pt}
\centerline{\hspace{-2.1cm}\raisebox{1.3cm}[0pt][0pt]{$\dot I_\Lambda^{ph2}$}\hspace{3.1cm}\raisebox{0.9cm}[0pt][0pt]{$\dot I_\Lambda^{ph1}$}}
\caption{Negligible terms in the flow equation for $\partial_\Lambda\Gamma$. The grey circles stand for any sub-diagram.}\label{neglected}
\end{figure}

The same kind of differential equations are obtained in an analogous way for the p-h channels\footnote{Remember that the formal products of functions mean something different in Eq.~\eref{rg2} than in Eq.~\eref{rg1}. The formal product in the p-p channel is defined by Eqs.~\eref{formalpp} and \eref{BSpp}, whereas the formal product in the p-h channels is defined by Eqs.~\eref{explicitph2} and \eref{formalph2}.}. Using Eqs.~\eref{scBS} and \eref{scseries}, we obtain
\begin{eqnarray}
 \dot R^{s,c}&=&-I^{s,c}\dot D^{ph}\Gamma^{s,c}-I^{s,c}D^{ph}\dot R^{s,c}\label{analog}\\
&=&-I^{s,c}\dot D^{ph}\Gamma^{s,c}+I^{s,c}D^{ph}I^{s,c}\dot D^{ph}\Gamma^{s,c}-\cdots\\
&=&-I^{s,c}\dot D^{ph}\Gamma^{s,c}-R^{s,c}\dot D^{ph}\Gamma^{s,c}\\
&=&-\Gamma^{s,c}\dot D^{ph}\Gamma^{s,c}.\label{rg2}
\end{eqnarray}
The only point to be verified is that the terms $\dot I^{s,c}D^{ph}\Gamma^{s,c}$ and $I^{s,c}D^{ph}\dot I^{s,c}$, which have been left out in Eq.~\eref{analog}, are in fact of subleading order in $\Lambda$. Writing out  $\dot I^{s,c}$ in terms of $\dot I^{ph1}$ and $\dot I^{ph2}$, one can verify that all the neglected terms are of the form shown graphically in Fig.~\ref{neglected}. They can be shown to be negligible in analogy with the example of Appendix~\ref{parquetex}. Equation \eref{rg2}, expressed in terms of $R^{ph1}$ and $R^{ph2}$ leads to
\begin{eqnarray}
\dot R^{ph1}&=&X\left((X\Gamma)\dot D^{ph}(X\Gamma)\right)\\
\dot R^{ph2}&=&-2\,\Gamma\dot D^{ph}\Gamma+(X\Gamma)\dot D^{ph}\Gamma+\Gamma\dot D^{ph}(X\Gamma),
\end{eqnarray}
or written out:
\begin{eqnarray}
\dot R^{ph1}_\Lambda(k_1,k_2,k_3)&=&\frac1{\beta V}\sum_p\dot D^{ph}_{\Lambda,q_1}(p)\,
\Gamma_\Lambda(p,k_2,p+q_1)\Gamma_\Lambda(k_1,p+q_1,k_3)\label{RGph1}\\
\dot R^{ph2}_\Lambda(k_1,k_2,k_3)&=&\frac1{\beta V}\sum_p\dot D^{ph}_{\Lambda,q_2}(p)\,\left[-2\Gamma_\Lambda(k_1,p,p+q_2) \Gamma_\Lambda(p+q_2,k_2,k_3)\right.\label{RGph2}\\ 
 & &\!\!\!\!\!\left.+\Gamma_\Lambda(p,k_1,p+q_2) \Gamma_\Lambda(p+q_2,k_2,k_3)+\Gamma_\Lambda(k_1,p,p+q_2) \Gamma_\Lambda(k_2,p+q_2,k_3)\right],\nonumber
\end{eqnarray}
where $q_1=k_3-k_1$, $q_2=k_3-k_2$ are the direct and exchanged transfered momenta and $D_\Lambda^{ph}(p,q)=G_\Lambda(p)G_\Lambda(p+q)$.

Equations \eref{master}, \eref{RGpp}, \eref{RGph1} and \eref{RGph2} are the one-loop RG equations we were looking for. They describe the behavior of the vertex under a differential change of the energy scale $\Lambda$. For the sake of clarity, these equations are collected as follows:
\begin{equation}
\dot\Gamma(k_1,k_2,k_3)=\dot R^{pp}(k_1,k_2,k_3)+\dot R^{ph1}(k_1,k_2,k_3)+\dot R^{ph2}(k_1,k_2,k_3)\label{1loop}
\end{equation}
\begin{align*}
\dot R^{pp}_\Lambda(k_1,k_2,k_3)&=\frac1{\beta V}\sum_p\dot D^{pp}_{\Lambda,k}(p)\,\Gamma_\Lambda(k_1,k_2,k-p)\Gamma_\Lambda(p,k-p,k_3)\\
\dot R^{ph1}_\Lambda(k_1,k_2,k_3)&=\frac1{\beta V}\sum_p\dot D^{ph}_{\Lambda,q_1}(p)\,
\Gamma_\Lambda(p,k_2,p+q_1)\Gamma_\Lambda(k_1,p+q_1,k_3)
\end{align*}
\begin{eqnarray*}
\dot R^{ph2}_\Lambda(k_1,k_2,k_3)&=&\frac1{\beta V}\sum_p\dot D^{ph}_{\Lambda,q_2}(p)\,\left[-2\Gamma_\Lambda(k_1,p,p+q_2) \Gamma_\Lambda(p+q_2,k_2,k_3)\right.\\ 
 & &\!\!\!\!\!\!\left.+\Gamma_\Lambda(p,k_1,p+q_2) \Gamma_\Lambda(p+q_2,k_2,k_3)+
\Gamma_\Lambda(k_1,p,p+q_2) \Gamma_\Lambda(k_2,p+q_2,k_3)\right],
\end{eqnarray*}
where $k=k_1+k_2$ is the total frequency-momentum and $q_1=k_3-k_1$, $q_2=k_3-k_2$ are the direct and exchanged transfered frequency-momenta, respectively. The p-p and p-h propagators are defined as $D_{\Lambda,k}^{pp}(p)=G_\Lambda(p)G_\Lambda(k-p)$ and $D_{\Lambda,q}^{ph}(p)=G_\Lambda(p)G_\Lambda(p+q)$. 

These equations have a very simple diagrammatic interpretation. The three terms $\dot R^{pp}$, $\dot R^{ph1}$ and $\dot R^{ph2}$ correspond to the one-loop diagrams shown in Fig.~\ref{diags}, where the wavy lines now represent full vertices $\Gamma_\Lambda$ and the product of two single particle propagators has to be derived with respect to $\Lambda$. Since $\dot D^{pp/ph}_{\Lambda,q}(p)=\dot G_\Lambda(p)G_\Lambda(q\pm p)+G_\Lambda(p)\dot G_\Lambda(q\pm p)$, each diagram can be viewed as the sum of two terms, where one of the two lines represents the propagator $G_\Lambda(p)$ and the other is $\dot G_\Lambda(p)$, the propagator restricted to the energy scale $\Lambda$. 

Although we have defined $\Lambda$ as a sharp infrared cutoff in the band energy, the RG equations are also correct for a cutoff in the frequency or for smooth cutoffs. It is straightforward to use the finite temperature instead of a cutoff to regularize the theory. The RG equations for $\partial_T\Gamma_T$ are then obtained by replacing in Eq.~\eref{1loop}
\begin{equation}
\frac1{\beta V}\sum_p\dot D^\diamond(p,q)\ldots\ \to\frac1{V}\sum_{n\in\Z,\,\v p} \partial_T\left(\frac1\beta D^\diamond(p,q)|_{p_0=2\pi T(2n+1)}\right)\ldots,
\end{equation}
where $\diamond=pp,ph$. 

Different regularizations of the theory give the same result to leading logarithmic order. For instance, the vertex at zero temperature and finite infrared cutoff $\Lambda$ is, within logarithmic precision, equal to the vertex at temperature $T=\Lambda$ without cutoff.

Let us close this section with the remark that the present RG formalism allows the calculation, to leading order in the infrared cutoff, of various generalized susceptibilities, i.e. the linear response of the system to weak external perturbations. The required relations are derived in Appendix~\ref{corr}.

\section{Discussion of the low-energy behavior}\label{Discussion}

The leading terms  for $\Lambda\to0$ of the right-hand side of the one-loop RG equations behave like $\sim\Gamma_\Lambda^2\Lambda^{-1}$ (note that $\Lambda^{-1}$ is the derivative of the logarithm). However, the one-loop RG equations contain in general many non-leading terms, like $\dot R^{pp}_\Lambda$ if $\v k$ is not a p-p nesting vector and $\dot R^{ph1}_\Lambda$ if $\v q_1$ is not a p-h nesting vector, etc. These terms are diverging at most as $\sim \Gamma_\Lambda^2\log\Lambda$. 

Since the one-loop RG equations are only correct to leading order in $\Lambda$, only the leading terms are to be taken seriously. Taking into account subleading terms in the one-loop RG equations is inconsistent and arbitrary, since these subleading terms are comparable to terms that have been neglected. The contributions to the exact RG flow can be classified into three classes: 1. leading terms of the one-loop equation, 2. subleading terms of the one-loop equation and 3. terms which are neglected in the one-loop equation. One has to compare the subleading terms $\sim\Gamma_\Lambda^2\log\!\Lambda$ with the neglected terms.  According to a careful analysis by Salmhofer and Honerkamp \cite{Salmhofer01}, the first neglected terms are or the order $\sim\Gamma^3_\Lambda\log^2\!\Lambda$ (see Eq. 108 of \cite{Salmhofer01}). Schematically,
\begin{equation}
\dot\Gamma_\Lambda=\underbrace{\mbox{``leading terms''}}_{\sim\Gamma_\Lambda^2\Lambda^{-1}}+ \underbrace{\mbox{``subleading terms''}}_{\sim\Gamma_\Lambda^2\log\!\Lambda}+\underbrace{\mbox{``neglected terms''}}_{\sim\Gamma^3_\Lambda\log^2\!\Lambda}.
\end{equation}

The RG flow is split into three different energy regimes. 
\begin{enumerate}
\item {\bf High energies}, in which $\Lambda$ is not small. In this regime, the subleading terms of the one-loop equation dominate the neglected two-loop terms, provided $\Gamma_\Lambda\log\!\Lambda$ is small. Thus the one-loop RG equations can in principle be used to calculate $\Gamma_\Lambda$, starting with a cutoff equal to the bandwidth, where the vertex is given by the bare coupling. In this regime, there is no small parameter to further simplify the functional RG equations, so using them is technically very difficult. But in the high energy regime ($g\log\Lambda\ll1$) na\"\i ve perturbation theory provides a controlled and much more feasible method of calculation than the RG. 
\item {\bf Small energies}, where $\Lambda$ is small such that $(\Lambda\log\Lambda)^{-1}\gg\Gamma_\Lambda\log\!\Lambda\sim1$. Na\"\i ve perturbation theory breaks down in this regime, but the leading terms of the one-loop RG equations are superior to the neglected terms even if $\Gamma$ is not necessarily small. This is where the one-loop RG is most useful. Note, that the subleading terms are no longer superior to the neglected terms in this regime. Thus, they have also to be neglected to be consistent. The small parameter $\Lambda$ justifies to replace every vertex $\Gamma(k_1,k_2,k_3)$ in the one-loop RG equations by its value on the Fermi surface ($k_{0,1}=k_{0,2}=k_{0,3}=\xi_{\v k_1}=\xi_{\v k_2}=\xi_{\v k_3}=0$), for reasons given in the end of Section~\ref{ladder} (for the case where the instabilities only occur in the p-p channel) and in Appendix~\ref{parquetex} (for the nested case). 
\item {\bf The critical regime}, where $\Lambda$ is close to the critical energy scale $\Lambda_c$ at which $\Gamma_\Lambda$ is diverging. If $\Lambda$ is too close to $\Lambda_c$, the neglected terms are no longer negligible and the one-loop approximation is no longer accurate. Note however that the weaker the initial interaction is, the more $\Lambda$ can approach $\Lambda_c$ before the one-loop approximation breaks down. 
\end{enumerate}

This remark about subleading terms concerns in particular the self-energy corrections to the propagator $G_\Lambda$. As already pointed out after Eq.~\eref{seriespp}, the self-energy changes the shape and location of the Fermi surface. This effect is by no means negligible in perturbation theory. Other effects, such as the renormalization of the Fermi velocity or the reduction of the quasi-particle weight are of subleading order. So these effects cannot be taken into account consistently within the one-loop approximation. For this reason, it is a consistent approximation to replace the dressed electron propagators  $G_\Lambda$ in the one-loop RG equations by the bare ones $C_\Lambda$.

\section{Relation to the Wilsonian approach}\label{Wilson}

A key ingredient to the Wilsonian RG is the idea to replace the given problem by a different one with less degrees of freedom but with the same low energy behavior. To achieve this, the effect of the eliminated degrees of freedom is incorporated in a renormalization of the parameters of the effective low-energy theory. This strategy has been successfully followed using Brillouin-Wigner perturbation theory in the strong coupling limit of various many-body problems (see for example \cite{Emery,Auerbach,Fazekas}). The best known example is the Hubbard model at half-filling which is represented, in the limit of strong coupling, by a Heisenberg spin Hamiltonian in the limit of strong coupling. In the weak coupling limit, the application of the Brillouin-Wigner formalism to compute an effective Hamiltonian is less evident and has not been followed to our knowledge. A tractable implementation of the same idea (i.e. elimination of high energy degrees of freedom and renormalization of parameters of the effective theory) is given more easily in the functional integral representation \footnote{It should nevertheless be mentioned that the idea of effective Hamiltonians on a reduced Hilbert space led to a most powerful numerical tool for one-dimensional systems: the density matrix renormalization group (see \cite{Noack99} for an introduction). An alternative route to effective Hamiltonians was presented recently by Wegner \cite{Wegner94} and applied to the 2D Hubbard model \cite{Hankevych03} and other correlated systems \cite{Kehrein01,Heidbrink02}.}. Some basic definitions and results of the functional integral formalism are given in Appendix~\ref{grassmann}, for a detailed presentation, see \cite{Negele}. 

The bare propagator is endowed with an infrared cutoff $\Lambda$ on the band energy $C_\Lambda(k)=\Theta(|\xi_{\v k}|-\Lambda)C(k)$. The {\it effective interaction} depending on the energy scale $\Lambda$ is defined as
\begin{equation}
{\cal W}_\Lambda[\chi]=-\log  \int\!\ud\mu_{C_\Lambda}[\psi]\,
e^{-W[\psi+\chi]}\label{effint},
\end{equation}
where $\ud\mu_{C_\Lambda}[\psi]$ is the normalized Gaussian measure according to the propagator $C_\Lambda$ (Eq.~\eref{Gaussian}), and $W$ is the model interaction in terms of Grassmann variables Eq.~\eref{int}. 
Note that the integration with respect to $d\mu_{C_\Lambda}[\psi]$ is perfectly
 defined, 
although $C_\Lambda^{-1}$ is not. This can be seen most easily in the 
expansion of ${\cal W}_\Lambda[\chi]$ in terms 
of Feynman diagrams. The evaluation of these diagrams involves only 
$C_\Lambda$ and never $C_\Lambda^{-1}$. Whenever $C_\Lambda^{-1}$ 
appears in an intermediate step of a calculation, it may be 
regularized by replacing the zero in the Heavyside function by an 
infinitesimal number.

The effective interaction ${\cal W}_\Lambda[\chi]$, which depends on the Grassmann field $\chi$, has a twofold interpretation. On the one hand, we 
can restrict the field $\chi$ to the low energy degrees of freedom 
$\psi^{\rm <}_{k\sigma}=\Theta(\Lambda-|\xi_{\v k}|)\psi_{k\sigma}$. 
The object
\begin{equation}
S^{\rm eff}_\Lambda[\psi^{\rm <}]=(\bar\psi^{\rm <},C^{-1} \psi^{\rm <})-
{\cal W}_\Lambda
[\psi^{\rm <}]
\end{equation}
corresponds then to Wilson's effective action, which describes the system in 
terms of $\psi^{\rm <}$ only. In fact, for observables (or Green's functions) which depend only on the low energy fields, one can show
\begin{eqnarray}
\langle O[\psi^<]\rangle&=&\frac1Z\int{\cal D}\psi^<\int{\cal D}\psi^>\ e^{S[\psi^{<},\psi^{>}]}\ O[\psi^<]\\
&=&\frac1Z\int{\cal D}\psi^<\  e^{S^{\rm eff}_\Lambda[\psi^{<}]}\ O[\psi^<]
\end{eqnarray}
and
\begin{equation}
Z=\int{\cal D}\psi^<\int{\cal D}\psi^>\ e^{S[\psi^{<},\psi^{>}]}=\int{\cal D}\psi^<\  e^{S^{\rm eff}_\Lambda[\psi^{<}]}.
\end{equation}
On the other hand, ${\cal W}_\Lambda$ is the generating functional of amputated
 connected correlation functions with infrared cutoff $\Lambda$ because of the
 identity  
\begin{equation}
\log Z_\Lambda[\eta]=-(\bar\eta,C_\Lambda \eta)-{\cal W}_\Lambda[C_\Lambda\eta]
,\label{gen}
\end{equation}
where $Z_\Lambda$ is given by Eq.~\eref{Zeta}, with $C$ replaced by 
$C_\Lambda$. To prove Eq.~\eref{gen}, we write out the Gaussian measure in (Eq.~\eref{Gaussian}) explicitly. Then, only in the numerator, we perform a shift of the integration variables $\psi\to \psi-\chi$. Equation \eref{gen} follows finally by putting $\chi=C_\Lambda\eta$ and taking the logarithm.    

As a consequence of Eq.~\eref{gen}, the quadratic part of ${\cal W}_\Lambda$ is related
 to the self-energy $\Sigma_\Lambda$ by 
\begin{eqnarray}
\left.\frac{\delta^2}{\delta\!\chi_{\sigma k}\delta\!\bar\chi_{\sigma k}}
{\cal W}_\Lambda[\chi]\right|_{\chi=0}&=&-C^{-1}_\Lambda(k)-\langle
\psi_{\sigma k}\bar\psi_{\sigma k}\rangle_{\rm \Lambda}\,C^{-2}_\Lambda(k)\\
&=&\frac{\Sigma_\Lambda(k)}{1-C_\Lambda(k)\Sigma_\Lambda(k)},\label{self}
\end{eqnarray}
where we have used the following identity for the full electron propagator 
$G_\Lambda(k)=-\langle\psi_{\sigma k}\bar\psi_{\sigma k}\rangle_{\rm \Lambda}
=C_\Lambda(k)(1-C_\Lambda(k)\Sigma_\Lambda(k))^{-1}$. 
Therefore in the case $|\xi_{\v k}|<\Lambda$ the right-hand side of Eq. 
\eref{self} simply becomes $\Sigma_\Lambda(k)$. 

Similarly the quartic part of  ${\cal W}_\Lambda$ is related to the one particle irreducible vertex $\Gamma_\Lambda$. In fact,
differentiating  Eq.~\eref{gen} we find 
\begin{eqnarray}
\left.\frac{\delta^{4}{\cal W}_\Lambda[\chi]}
{\delta\!\chi_{\sigma k_4}\delta\!\chi_{\sigma' k_3}\delta\!\bar
\chi_{\sigma' k_2}\delta\!\bar\chi_{\sigma k_1}}\right|_{\chi=0}
&=&-\langle\psi_{\sigma k_1}\psi_{\sigma' k_2}\bar\psi_{\sigma' k_3}\bar
\psi_{\sigma k_4}\rangle_{\rm c,\Lambda}\prod_{i=1}^4C_\Lambda^{-1}(k_i)\\
&=&-\frac{\Gamma_\Lambda^{\sigma\sigma'}(k_1,\ldots,k_4)}{\beta V\prod_{i=1}^4
[1-C_\Lambda(k_i)\Sigma_\Lambda(k_i)]}.\label{vert}
\end{eqnarray}
In the last line we have used Eq.~\eref{G2} and the definition of the vertex (Eq.~\eref{vertexdef}).

The quartic part of ${\cal W}_\Lambda$ is of the same form as Eq.~\eref{int} 
with an effective coupling function $g_\Lambda(k_1,k_2,k_3)$ that now depends on the frequencies as well as the momenta. Taking functional derivatives of Eq.~\eref{int}, we find for $|\xi_{\v k_i}|<\Lambda$,  
\begin{equation}
(1-\delta_{\sigma \sigma'}X)g_\Lambda(k_1,k_2,k_3)=-\Gamma^{\sigma \sigma'}_{\Lambda}(k_1,k_2,k_3)
\end{equation}
and thus
\begin{equation}
g_\Lambda(k_1,k_2,k_3)=-\Gamma_{\Lambda}(k_1,k_2,k_3).\label{correspondence}
\end{equation}
We summarize that $g_\Lambda$ is equal, up to the sign, to a connected amputated 
correlation function if all $|\xi_{\v k_i}|>\Lambda$ and to the one-particle irreducible vertex in the opposite case. $g_\Lambda$ is therefore not continuous at $|\xi_{\v k_i}|=\Lambda$.
A formal and non-perturbative proof of these relations was given  by Morris 
\cite{Morris94} for a bosonic field theory. The derivation given above is perturbative, but a generalization of the non-perturbative proof of Morris to fermions appears to be straightforward. Morris has also shown that $\Sigma_\Lambda$ and 
$\Gamma_\Lambda$ are continuous at $|\xi_{\v k_i}|=\Lambda$, in contrast to $g_\Lambda$.

\section{Other one-loop RG equations}\label{compare}

The effective interaction satisfies the following exact RG equation for $\partial_\Lambda{\cal W}_\Lambda=\dot{\cal W}_\Lambda$:
\begin{equation} 
\dot{\cal W}_\Lambda[\chi]=\sum_{\sigma,k}\dot C_\Lambda(k)\
\frac{\delta^2{\cal W}_\Lambda[\chi]}{\delta\!\chi_{\sigma k}\delta\!
\bar\chi_{\sigma k}}-\sum_{\sigma,k}\dot C_\Lambda(k)\ \frac{\delta{
\cal W}_\Lambda[\chi]}{\delta\!\chi_{\sigma k}}\frac{\delta{
\cal W}_\Lambda[\chi]}{\delta\!\bar\chi_{\sigma k}}.\label{exact}
\end{equation} 
This equation was first derived by Polchinski in the context of a scalar field theory \cite{Polchinski84}. Zanchi and Schulz \cite{Zanchi,Zanchi00} proposed to develop ${\cal W}_\Lambda$ up to order six in the fermionic variables and to neglect terms of higher order. Terms of order six are not present in the original interaction but they are produced by the RG procedure. Their effect is then to renormalize the effective coupling function $g_\Lambda$. The result is a closed one-loop equation for the coupling function 
$g_\Lambda(k_1,\ldots,k_4)$  where $|\xi_{\v k_i}|<\Lambda$. It is identical, within the correspondence $g_\Lambda=-\Gamma_\Lambda$, to our Eq.~\eref{1loop} with one difference. In the RG equation of Zanchi and Schulz, the vertices on the right hand-side are not evaluated at the scale $\Lambda$, but at a higher scale $\tilde\Lambda$, which is given by the band energy of the single-particle propagators, i.e. $\tilde\Lambda=\Max\{|\xi_{\v p}|,|\xi_{\v k-\v p}|\}$ in the p-p term of Eq.~\eref{1loop}, $\tilde\Lambda=\Max\{|\xi_{\v p}|,|\xi_{\v p+\v q_{1}}|\}$ in the p-h 1 term and $\tilde\Lambda=\Max\{|\xi_{\v p}|,|\xi_{\v p+\v q_{2}}|\}$ in the p-h 2 term. Their equation is thus non-local in $\Lambda$. It is a flow equation with memory, i.e. the flow at scale $\Lambda$ doesn't only depend on the vertex function $\Gamma_\Lambda$ but also on the history of the flow.

Since this is not very convenient, it was proposed \cite{Halboth00} 
to develop Eq.~\eref{exact} into Wick ordered polynomials of the fermionic 
variables instead of monomials as it was done above. Wick ordering with
 respect to the low energy propagator $D_\Lambda=C-C_\Lambda$ 
results in the same one-loop equation as above but now all the couplings are 
evaluated at the actual RG variable $\Lambda$ and $-\ud\left[C_\Lambda(p)
C_\Lambda(q)\right]/{\ud\Lambda}$ has to be replaced by $\ud\left[D_\Lambda(p)
D_\Lambda(q)\right]/{\ud\Lambda}$, i.e., one propagator is at the energy $\Lambda$ and the energy of the second propagator is now restricted to be {\it smaller} than $\Lambda$. This is different from our one-loop RG equation, where the second propagator is restricted to higher energies. 

We have seen above that the coupling function of the low-energy effective theory equals, up to a sign, the one-particle irreducible vertex $\Gamma_\Lambda$. This correspondence can not be generalized to higher order vertices, i.e. it would be completely wrong to say that the sixth order term of ${\cal W}_\Lambda$ is related to the one-particle irreducible part of the three particle Green's function and so forth. In addition, the correspondence Eq.~\eref{correspondence} relies on the choice of the cutoff which ensures that $C_\Lambda(p)=0$, for $|\xi_{\v p}|<\Lambda$. The correspondence  no longer holds for an alternative scheme, where for example the finite temperature is used to regularize the theory instead of the infrared cutoff. 

The general one-particle irreducible vertices are obtained by performing a Legendre transformation on the functional ${\cal W}_\Lambda$. Wetterich has presented a renormalization group scheme for bosonic field theories, working with this Legendre transformed quantity rather that with the effective action defined in Eq.~\eref{effint}  \cite{Wetterich93,Tetradis94}. The idea was implemented recently for the many-fermion problem in \cite{HSFR01,Salmhofer01}. The resulting RG equation is identical to Eq.~\eref{1loop} apart from self-energy terms  which should be, as we argued above, neglected for consistency reasons. In fact, if the cutoff is introduced by some function multiplying the bare propagator
\begin{equation}
C_\Lambda(p)=\frac{\theta_\Lambda(p)}{ip_0-\xi_{\v p}},
\end{equation}
then the full propagator is given by
\begin{eqnarray}
G_\Lambda(p)&=&\frac{C_\Lambda(p)}{1-C_\Lambda(p)\Sigma_\Lambda(p)}\\
&=&\frac{\theta_\Lambda(p)}{ip_0-\xi_{\v p}-\theta_\Lambda(p)\Sigma_\Lambda(p)}.
\end{eqnarray}
The derivative with respect to $\Lambda$ gives
\begin{equation}
\dot G_\Lambda(p)=\frac{\dot\theta_\Lambda(p)(ip_0-\xi_{\v p})}{\left(ip_0-\xi_{\v p}-\theta_\Lambda(p)\Sigma_\Lambda(p)\right)^2}+\frac{\theta^2_\Lambda(p)\dot\Sigma_\Lambda(p)}{(ip_0-\xi_{\v p}-\theta_\Lambda(p)\Sigma_\Lambda(p))^2}.\label{Gdot}
\end{equation}
The first term of the right hand-side is the single-scale propagator of Ref.~\cite{HSFR01}. The one-loop RG equations of Salmhofer and Honerkamp are obtained from our Eq.~\eref{1loop} by omitting the second term of Eq.~\eref{Gdot}.

The RG equations of the three groups \cite{Zanchi00,Halboth00,HSFR01} differ in the treatment of the diagrams which give not a leading order contribution in the limit $\Lambda\to0$. For example, the (leading order) p-p diagram at total momentum $k=0$ features two internal propagators with exactly opposite momenta. They have exactly the same band energy. So the non-locality of the Zanchi-Schulz equation is not present in this diagram. For the same reason, the energy-constraint of the Wick-ordered scheme can not have any effect in this case. So the three RG schemes treat such diagrams identically. The same is true for the p-h diagrams in the case of perfect nesting. However, as we have argued in Section~\ref{floweq}, only the leading order terms of these RG equations make sense. The subleading terms are of the same order as others that have been neglected. In fact, a consistent treatment of subleading terms requires going beyond the one-loop approximation.

The equivalence of the different RG equations to leading order can also be understood in the following way. By successively integrating the one-loop RG equations (Eq.~\eref{1loop}) and expressing the result in terms of  $\Gamma_{\Lambda_0}\approx-g$, one obtains the full series of parquet diagrams. However the structure of these RG equations introduces a constraint on the energies of internal lines. For example the parquet diagram of Fig.~\ref{parquetfig2} is generated by the one-loop RG with the following constraint on the propagators 1,2,3 and 4
\begin{equation}
\Min\{|\xi_1|,|\xi_2|\}\leq\Min\{|\xi_3|,|\xi_4|\}.\label{constraint}
\end{equation} 
Higher order parquet diagrams are generated with similar constraints, i.e. with a certain ``ordering'' of the band energies, when one goes from the inner loops to the exterior loops of a given parquet diagram. It can be checked by introducing this constraint into Eq.~\eref{2loop} of Appendix~\ref{parquetex}, that the constraint does not change the value of the diagram to leading logarithmic order in $\Lambda$. 

\begin{figure}  
\centerline{\includegraphics[width=10cm]{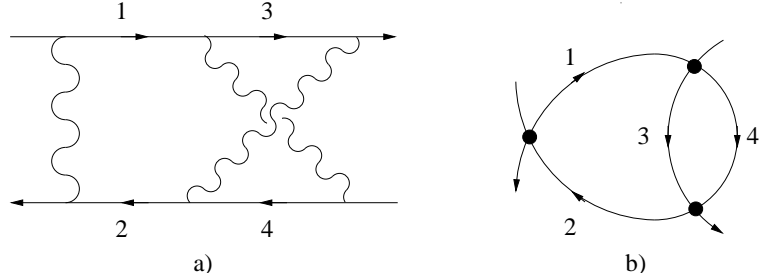}}
\caption{Fig. a) shows an example of a two loop parquet diagram, where the internal electron propagators are numbered from 1 to 4. Fig. b) shows the same diagram, but the wavy interaction lines are contracted to points in order to make 
the loop structure more transparent.}\label{parquetfig2}
\end{figure}

The different RG equations \cite{Zanchi00,Halboth00,HSFR01} all generate the whole series of parquet diagrams, but the constraints are different. In the Wick ordered scheme, the constraint \eref{constraint} is changed into 
\begin{equation}
\Max\{|\xi_1|,|\xi_2|\}\leq\Max\{|\xi_3|,|\xi_4|\}.\nonumber
\end{equation}
The RG equation of Zanchi and Schulz introduces the most restrictive constraint
\begin{equation}
\Max\{|\xi_1|,|\xi_2|\}\leq \Min\{|\xi_3|,|\xi_4|\},\nonumber
\end{equation}
i.e. both propagators of the inner loop are higher in energy than both propagators of the exterior loop. All these constraints are irrelevant for the leading logarithmic order in $\Lambda$. We conclude that our one-loop RG equations and those of Refs.~\cite{Zanchi00,Halboth00,HSFR01} are all equivalent to leading  order in $\Lambda$. 

\section{Conclusion}\label{Conclusion}

In conclusion, we have presented the RG method for weakly interacting electrons to leading (one-loop) order using a simple framework. We have chosen the point of view of partial summations of the perturbation series, taking into account the leading infrared-diverging terms and neglecting subleading contributions in an unbiased and consistent way. The procedure is controlled by two small parameters, the bare interaction strength and the energy scale of interest. 

In the case of a generic Fermi surface in more than one dimension, this procedure leads to the ladder approximation. Our calculations for the two-dimensional Hubbard model away from half filling show a pole in the scattering vertex $\Gamma(p_1,p_2,p_3)$ at $p_1+p_2=0$, where $p_{1,2}$ are the frequency-momenta of the incoming particles. The standard interpretation of this pole is the development of a two-particle bound state, i.e. of a pairing instability. The symmetry of the pairing state is $d$-wave. The pairing mechanism can be interpreted in terms of an effective attraction between electrons which is created by spin fluctuations. The critical energy scale at which the pole is created increases drastically as half-filling is approached. The reasons are twofold. First, antiferromagnetic spin fluctuations are enhanced due to approximate nesting and second, the density of states increases due to the proximity of a Van Hove singularity.

 However, if the Fermi surface is nested, the summation of p-p ladder diagrams is no longer sufficient, but the more general RG formalism involving p-h as well as p-p contributions is required.  As opposed to some more modern formulations \cite{Shankar94,Zanchi00,Halboth00,HSFR01}, we have derived the one-loop RG equations by identifying the leading logarithmic terms. We feel that the less modern route chosen here makes the meaning of the one-loop approximation more transparent. In particular, it gives a natural guideline for deciding which terms are to be included in the calculation and which ones are to be neglected for reasons of consistency. For example, we argue in Section~\ref{Discussion} that if the Fermi surface is not nested and the Fermi level is not close to a Van Hove singularity, the one-loop RG is not better than the ladder approximation. This is also true in the presence of Umklapp scattering, in contrast to the assertions of Ref.~\cite{HSFR01}. Furthermore, in Section~\ref{compare}, we compare  three different versions of RG equations, which were discussed in the literature. We find that all of these approaches are equivalent within the regime of applicability of the one-loop approximation. 

 The RG approach is most useful in one dimension, or in the case of a flat Fermi surface with many nesting vectors (see Section~\ref{nesting} for a precise definition). If there is only one single nesting vector, the situation is simpler. Consider for example an anisotropic square lattice at half filling. The dispersion is given by $\xi_{\v k}=-2(t_x\cos k_x+t_y\cos k_y)$, where $0<|t_y/t_x|<1$ (not too close to neither 0 or 1). Because the Fermi surface is perfectly nested by $(\pi,\pi)$, there are infrared singularities in the p-h as well as in the p-p channel. However, the non-ladder parquet diagrams appear not to be of the leading order in this case. As a consequence one can just sum the p-h 1, p-h 2 and p-p ladders separately. In other words, the one-loop RG equations decouple (to leading order) into three different sets, related to pairing-, spin- and charge instabilities, respectively \cite{thesis}. The study of the p-h instabilities is then in perfect analogy to our treatment of superconductivity in Section~\ref{ladder} and \ref{superconductivity}. A repulsive Hubbard-like interaction will lead to a spin-density wave instability. 

 More difficult and also more interesting is the case, where the Fermi surface passes through saddle points of the dispersion relation. Because the dispersion is quadratic rather than linear at the saddle points, they render infrared divergences more dangerous, particularly in two dimensions, where saddle points lead to a logarithmic  Van Hove singularity in the density of states. 

We have recently studied the half-filled nearest-neighbor tight-binding band on a square lattice (i.e. the case mentioned above, but with $t_x=t_y$), where the Fermi surface is a perfect square and contains two saddle points. Our investigation yields a rich phase diagram as a function of the model interaction \cite{Binz02}. Apart from $s$- and $d_{x^2-y^2}$-wave superconductivity, spin- and charge-density waves as well as two phases with circulating charge or spin currents ($d$-density waves) appear as possible instabilities. Unlike numerical studies of the functional RG equations \cite{Zanchi00,Halboth00,Honerkamp01} which rely on a discretization, we do not obtain a simultaneous divergence of different, say superconducting and spin-density wave susceptibilities as the energy scale is lowered, but only of the leading susceptibility. Although there is no  decoupling of the RG equations into independent sets, it turns out that  the leading channel asymptotically decouples from the others as an instability is approached \cite{Binz02}.

The case of the  half-filled nearest-neighbor tight-binding band is special because it features perfect nesting and saddle points at the same time. Considerable effort has been spent on the more general case of a Fermi surface which is at Van Hove filling but not nested. This case is particularly interesting because it potentially contains new physics such as non Fermi liquid behavior \cite{Dzyaloshinskii96,IK01}, itinerant ferromagnetism \cite{IKK01,HSPRL01} or Pomeranchuk instabilities \cite{Halboth01}. Unfortunately, the one-loop RG equations are likely to be unreliable in this case. The reason is that the one-loop equation generates a correct summation merely of the {\it leading} divergent terms in the perturbation series, but the above-mentioned phenomena depend on  {\it subleading} logarithmic divergences. This can be seen by considering the following example.  To study the ferromagnetic (Stoner) instability, we have to calculate the scattering vertex $\Gamma(k_1,k_2,k_3)$ for an infinitely small momentum transfer, $\v k_3-\v k_1\to\v 0$.  The instability is usually created by p-h ladder diagrams (Fig.~\ref{ph1ladders}). In the case of Van Hove filling,  the zero-momentum limit of the p-h bubble\footnote{The p-h bubble is defined by $B^{ph}(q)=1/(\beta V)\sum_p C(p)C(p+q)$ and is equal up to the sign to the non-interacting spin susceptibility.}  diverges as $\sim\log T$ for low temperatures\footnote{It is not possible to study the long-wavelength p-h response in the presence of  an infrared momentum cutoff. The regularization has therefore to be done in another way, for example by a finite temperature.} as a consequence of the infinite density of states. This is a subleading divergence compared to the behavior of the zero-momentum p-p bubble, which is $\sim\log^2 T$ in the case of Van Hove filling. The third-order diagram of the   p-h ladder series behaves as $U^3\log^2T$ since it contains two low-momentum p-h bubbles, each contributing with one logarithm. The one-loop RG theory in addition generates parquet diagrams, where p-p and p-h bubbles are mixed. The diagram shown in Fig.~\ref{parquetfig2} contains one p-h bubble and one (internal) p-p bubble. If the external legs are chosen close to the saddle points, its leading behavior is $U^3\log^3T$. The error from the one-loop approximation is of subleading logarithmic order,  i.e. in our case of the order  $U^3\log^2T$. The problem is obviously that the error produced by the one-loop approximation is of the same order as the term which produces the ferromagnetic instability. The difficulty is not only encountered for third-order diagrams but proliferates to every order in perturbation theory.  We conclude that the one-loop RG is an uncontrolled approximation for discussing ferromagnetism.

A consistent weak-coupling treatment should at least include a self-consistent determination of the Fermi surface, self-energy corrections to both the quasi-particle weight and the Fermi velocity as well as two-loop corrections to the RG equations of the vertex. Although some progress has already been made on certain of these aspects \cite{Zanchi01,HS03,neumayr03}, a satisfactory control of the Van Hove case has not yet been obtained. New techniques will have to be developed to settle these issues.

\begin{acknowledgement}
B. B. thanks N. Dupuis, A. Ferraz, F. Gebhard, F. Guinea, G. I. Japaridze, A. P. Kampf, A. A. Katanin, W. Metzner, T. M. Rice, V. Rivasseau, A. Rosch,
B. Roulet, M. Salmhofer, G. S. Uhrig, F. Vistulo de Abreu, D. Vollhardt, M. A. H. Vozmediano, and D. Zanchi for useful and stimulating discussions. This work was supported by the Swiss National Foundation through grant no. 20-61470.00, the National  Centre of Competence in Research MaNEP (Materials with Novel Electronic Properties) and the Rectors' Conference of the Swiss Universities via a special mobility grant.
\end{acknowledgement}

\appendix
\section{Definition and properties of the vertex function}\label{Green}

In this appendix, we give the definition of the imaginary-time two-particle Green's function and the vertex function. These quantities are not directly observable but they are analytically related to various physical response functions \cite{Abrikosov,Fetter,Negele}. 

The two-particle Green's function is defined as
\begin{equation}
G^{\sigma\sigma'}(\tau_1,\tau_2,\tau_3,\v k_1,\v k_2,\v k_3)=V \langle\,Tc^{}_{\sigma \v k_1}(\tau_1)\,c^{}_{\sigma' \v k_2}(\tau_2)\,c^\dagger_{\sigma' \v k_3}(\tau_3)\,c^\dagger_{\sigma \v k_1+\v k_2-\v k_3}(0)\,\rangle
\end{equation}
where $V$ is the system volume (i.e. the number of lattice sites) and $T$ is the time ordering operator. The imaginary time dependence of any operator is given by  
\begin{equation}
O(\tau)=e^{\tau(H-\mu N)}\,O\,e^{-\tau(H-\mu N)},\label{imagHeis}
\end{equation}
where $\mu$ is the chemical potential and  $N=\sum_{\v k\sigma}c^\dagger_{\v k\sigma}c^{}_{\v k\sigma}$ the particle number operator. The Fourier transform with respect to the $\tau$ variables,
\begin{equation}
G^{\sigma\sigma'}(k_1,k_2,k_3)=\int_0^\beta\ud\tau_1\ud\tau_2\ud\tau_3\, e^{i(k_{01}\tau_1+k_{02}\tau_2-k_{03}\tau_3)}G^{\sigma\sigma'}(\tau_1,\tau_2,\tau_3,\v k_1,\v k_2,\v k_3),
\end{equation}
is a function of the $D+1$-dimensional frequency-momentum vectors $k_i=(k_{0i},\v k_i)$. The Matsubara frequencies $k_{0i}$ run over the odd multiples of $\pi/\beta$, because $G^{\sigma\sigma'}$ is anti-periodic with period $\beta$ in each of the $\tau$ variables. 

The vertex function $\Gamma^{\sigma\sigma'}(k_1,k_2,k_3)$, defined as the one-particle irreducible part of $G^{\sigma\sigma'}(k_1,k_2,k_3)$, satisfies the relation
\begin{equation}
G^{\sigma\sigma'}(k_1,k_2,k_3)=\beta V(\delta_{k_2,k_3}-\delta_{\sigma,\sigma'}\delta_{k_1,k_3})G(k_1)G(k_2)+\Gamma^{\sigma\sigma'}(k_1,k_2,k_3)\prod_{j=1}^4G(k_j),\label{vertexdef}
\end{equation}
where $G(k)$ is the exact one-particle Green's function and $k_4=k_1+k_2-k_3$. 

At this point it is worthwhile to investigate the symmetries of the vertex function. A simple $SU(2)$ transformation shows that for a spin-rotation invariant system,
\begin{equation}
\Gamma^{\up\up}=(1-X)\Gamma^{\up\down},
\end{equation}
where $X\Gamma(k_1,k_2,k_3)=\Gamma(k_2,k_1,k_3)$. 
Because the whole information is contained in the function $\Gamma^{\up\down}$, we will only consider the vertex of anti-parallel spins from now on and omit the spin indices (i.e. $\Gamma=\Gamma^{\up\down}$). The permutation symmetry gives 
\begin{equation}
\Gamma(k_1,k_2,k_3)=\Gamma(k_2,k_1,k_1+k_2-k_3).
\end{equation}
Time reversal invariance, spin rotation invariance and parity imply
\begin{equation}
\Gamma(k_1,k_2,k_3)=\Gamma(k_1+k_2-k_3,k_3,k_2). \label{tgamma}
\end{equation}
From the behavior under complex conjugation one obtains
\begin{equation}
\overline{\Gamma(k_1,k_2,k_3)}=\Gamma(\overline{k_1+k_2-k_3},\overline{k_3},\overline{k_2}),
\end{equation}
where $\overline{k}=(-k_0,\v k)$. It follows that $\Gamma$ is real if the frequencies are put to zero.

\section{Parquet diagrams: An example}\label{parquetex}

\begin{figure}  
\centerline{\includegraphics[width=6cm]{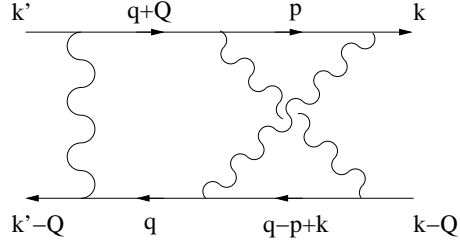}}
\caption{Example of a two loop parquet diagram.}\label{parquetfig}
\end{figure}

A typical parquet diagram, is shown in Fig.~\ref{parquetfig}. It is a p-p diagram embedded into a p-h 2. For simplicity, we choose the total momentum and the external frequencies $k_0$, $k'_0$ and $Q_0$ to be zero and we assume a constant coupling function $g$. In addition, we assume that $\v Q$ is a perfect nesting vector, such that $\xi_{\v q+\v Q}=-\xi_{\v q}$. The value of the diagram (at $T\to0$) is then
\begin{equation}
\frac{2g^3}{(\beta V)^2}\sum_{pq}C_\Lambda(q)C_\Lambda(q+Q)C_\Lambda(p)C_\Lambda(q-p+k).
\end{equation}
After integration with respect to $p_0$ and $q_0$, we obtain 
\begin{equation}
(74)=-\frac{g^3}{V^2}\sum_{\v p\v q}{ }^{'} \frac{\Theta(\xi_{\v p}\xi_{\v q-\v p+\v k})}{|\xi_{\v q}|(|\xi_{\v q}|+|\xi_{\v p}|+|\xi_{\v q-\v p+\v k}|)}=-\int_\Lambda^W\!\ud\xi\,\int_{2\Lambda}^W\!\ud\xi'\,\frac{\tilde\nu(\xi,\xi')}{|\xi|(|\xi|+|\xi'|)},
\end{equation}
 where the sum $\sum_{\v p\v q}'$ is over all momenta $\v p$ and $\v q$ allowed by the infrared cutoff 
 and
\begin{equation} 
\tilde\nu(\xi,\xi')=\frac1{V^2}\sum_{\v p\v q}\,\delta(\xi-\xi_{\v q})\delta(\xi'-\xi_{\v p}-\xi_{\v q-\v p+\v k})\Theta(\xi_{\v p}\xi_{\v q-\v p+\v k}). 
\end{equation}

If the Fermi level is not at a Van Hove singularity, $\tilde\nu(\xi,\xi')$ is usually bounded. Depending on the situation, $\tilde\nu$ could even be suppressed to zero for $\xi,\xi'\to0$. This would reduce the singularity of the diagram in such a way that it can be neglected within the leading order approximation. The cases where parquet diagrams are most singular are those where $\tilde\nu(\xi,\xi')$ is positive for $\xi,\xi'\to0$ (and can thus be approximated by a constant). It means that to leading order, our diagram is proportional to the following integral:
\begin{equation}
P(\Lambda)=\int_\Lambda^W\frac{\ud\xi}{\xi}\,\underbrace{\int_\Lambda^W\frac{\ud\xi'}{\xi+\xi'}}_{F_\Lambda(\xi)}=\frac12\log^2\frac W\Lambda.\label{2loop}
\end{equation}
For further convenience we have introduced the symbol $F_\Lambda(\xi)$ for the value of the internal p-p bubble. The result shows that indeed our diagram is of the leading order $g^3\log^2\Lambda$.

To obtain its contribution to the RG equation, we take the derivative with respect to $\Lambda$
\begin{equation}
-\partial_\Lambda P(\Lambda)=\frac1\Lambda F_\Lambda(\Lambda)+\int_\Lambda^W\!\frac{\ud\xi}{\xi(\xi+\Lambda)}\label{toyrg}
\end{equation}
The first term is obtained by putting the large loop variable $\xi$ to the energy scale $\Lambda$. The second term comes from the derivative of the internal p-p loop. A simple calculation leads us to two important observations:
\begin{enumerate}
\item The second term (equal to $\Lambda^{-1}\log\frac{2W}{W+\Lambda}\sim\Lambda^{-1}$) is negligible as compared to the first one ($\sim\Lambda^{-1}\log\Lambda$). This result can be generalized as follows. If a diagram is reducible in one channel, the dominant contributions to its $\Lambda$-derivative will be obtained by deriving only the propagators connecting the irreducible blocks and {\it not} the irreducible blocks themselves. This is exactly what we used in the derivation of the one-loop flow equations in Section~\ref{floweq}.
\item We can replace $F_\Lambda(\Lambda)$ in Eq.~\eref{toyrg} by $F_\Lambda(0)$ without changing the result to leading order. In fact, integrating the simplified RG equation $-\partial_\Lambda P(\Lambda)=\frac1\Lambda F_\Lambda(0)$, we obtain indeed the correct result for $P(\Lambda)$. This result can  be generalized as follows. In the RG equation, one can replace any vertex $\Gamma(k_1,k_2,k_3)$ by its value at $\xi_1=\xi_2=\xi_3=0$, i.e. the momenta $k_i$ can be projected onto the Fermi surface. In this sense the dependence of the vertex on the band energies $\xi_i$ is irrelevant. The same is true for its dependence on the frequencies. The replacement $F_\Lambda(\Lambda)\to F_\Lambda(0)$ is only justified in the differential RG equation Eq.~\eref{toyrg}, but {\it not} in the integral expression Eq.~\eref{2loop}, since $\int_\Lambda^W\ud\xi/\xi\, F_\Lambda(0)=\log^2(W/\Lambda)\neq P(\Lambda)$. It is important to note that although the dependence of the vertex on the band energies and frequencies is irrelevant in the sense explained above, this dependence does not need to be weak at all! In our example, $F_\Lambda(\xi)$ depends clearly on $\xi$.
\end{enumerate}

\section{Linear response}\label{corr}


In this appendix, we investigate the linear response of the system to a weak external perturbation. 

\subsection{Generalized susceptibilities for superconductivity}

Let the system be coupled to a time dependent, real valued field $ \lambda(\v r,t)$, via an additional term in the Hamiltonian
\begin{equation}
H_{{\rm ext},\,t}=-\sum_{\v r} \lambda(\v r,t)\left( \Delta_A^{}(\v r)+ \Delta_A^\dagger(\v r)\right),
\end{equation} 
where
\begin{equation}
 \Delta_A^{}(\v r)=\sum_{\v r'}  f_A(\v r-\v r')\,c_{\v r'\down}c_{\v r\up},\label{DSC}
\end{equation}
is the locally defined superconducting order parameter. The index $A$ stands for the internal wave function $  f_A(\v r-\v r')$ of the Cooper pair. This term can also be written in Fourier space as 
\begin{equation}
H_{{\rm ext},\,t}=-V\sum_{\v k}\left(\lambda(-\v k,t)\Delta_A^{}(\v k)+\mbox{ h. c. }\right),
\end{equation}
where 
\begin{equation}
\Delta_A^{}(\v k)=\sum_{\v r}e^{-i\v k\v r}\Delta_A^{}(\v r)=\sum_{\v p}f_A(\v p)\,c_{\v p\down}c_{\v k-\v p\up}
\end{equation}
and the functions $\lambda(\v k)$ and $f_A(\v p)$ are the Fourier transforms\footnote{We use the conventions $F(\v r)=\frac1{V}\sum_{\v k}e^{i\v k\v r}F(\v r)$ and $F(\v k)=\sum_{\v r}e^{-i\v k\v r}F(\v r)$ for any quantity $F(\v r)$. The only exceptions are the creation, and annihilation operators. There, we use a pre-factor $V^{-1/2}$ for both transformations. Note that $F^\dagger(\v k)$ denotes the Hermitian conjugate of $F(\v k)$ rather than the usual Fourier transform of $F^\dagger(\v r)$. The Fourier transform in (real) time is defined as $F(t)=\int\!\frac{\ud\omega}{2\pi}\,e^{-i\omega t}F(\omega)$ and $F(\omega)=\int\!\ud t\,e^{i\omega t}F(t)$.} of $\lambda(\v r)$ and $f_A(\v r)$.

The coupling to the field induces in general a non-zero value of $\langle \Delta_B^{}(\v r)\rangle_t$, where the internal wave function $f_B$ can be different from $f_A$. The linear response is given by the retarded response function or susceptibility $ \chi^{BCS}_{{\rm ret},BA}(\v r-\v r',t-t')$, via
\begin{equation}
\langle \Delta_B(\v r)\rangle_t=\int\!\ud t'\,\sum_{\v r'} \chi^{BCS}_{{\rm ret},BA}(\v r-\v r',t-t')\, \lambda(\v r',t')
\end{equation}
or, after Fourier transformation in space and time, 
\begin{equation}
\langle\Delta_B(\v k)\rangle_\omega=\chi^{BCS}_{{\rm ret},BA}(\v k,\omega)\,\lambda(\v k,\omega).
\end{equation}
The global superconducting order parameter is given by $\Delta(\v k=0)$. A more convenient quantity is the Matsubara response function given by
\begin{equation}
\chi^{BCS}_{BA}(\v k,\nu)=\frac1{V}\int_0^\beta\!\ud\tau\,e^{i\nu\tau}\left\langle\Delta^{}_B(\v k,\tau)\Delta^\dagger_A(\v k)\right\rangle,\label{chiBCS1}
\end{equation}
where the Bose-Matsubara frequency $\nu$ is restricted to even multiples of $2\pi/\beta$ and the imaginary time dependence of operators is given by Eq.~\eref{imagHeis}. The retarded response function is obtained by analytic continuation 
\begin{equation}
\chi(\nu)\ \MapRight{i\nu\to\omega+i\delta}\ \chi_{\rm ret}(\omega),
\end{equation}
where $\delta$ is an infinitesimally positive (dissipative) term.

Equation \eref{chiBCS1} together with Eq.~\eref{vertexdef}, give 
\begin{eqnarray}
\chi^{BCS}_{BA}(k)&=&\frac1{V}\int_0^\beta\!\ud\tau\,e^{i\nu\tau}\sum_{\v p,\v p'}f_B(\v p)\overline{f_A(\v p')}\,\left\langle c^{}_{\v p\down}(\tau)c^{}_{\v k-\v p\up}(\tau)c^\dagger_{\v k-\v p'\up}c^\dagger_{\v p'\down}\right\rangle\\
&=&\frac1{(\beta V)^2}\sum_{p,p'}f_B(\v p)\overline{f_A(\v p')}\ G^{\down\up}(p,k-p,k-p'),\\
&=&\frac1{(\beta V)^2}\sum_{p,p'}f_B(\v p)\left[\beta V\delta_{p,p'}+D^{pp}_{k}(p)\,\Gamma^{BCS}_k(p,p')\right]D^{pp}_{k}(p')\,\overline{f_A(\v p')}\label{chiBCS},
\end{eqnarray}
where $k=(\nu,\v k)$,  $\Gamma^{BCS}_k(p,p')=\Gamma(p,k-p,k-p')$ and $\overline{f_A(\v p')}$ is the complex conjugate of $f_A(\v p')$.

It is argued at the end of Section~\ref{ladder} that the low energy behavior in the weak coupling limit is determined by the degrees of freedom close to the Fermi surface. Therefore the dependence of the form factors on $\xi$ is irrelevant. Let us perform the change of variables $\v p\,\to\,\xi_{\v p},\theta_{\v p}$, where $\xi$ is the band energy and $\theta$ is a suitable (angular) variable. The most elementary superconducting susceptibility is a function of two angular variables, which is obtained by putting $f_B(\v p)=\delta(\theta_{\v p}-\theta)$ and $f_A(\v p')=\delta(\theta_{\v p'}-\theta')$ in Eq.~\eref{chiBCS}, 
\begin{equation}
\chi^{BCS}(k,\theta,\theta')=\int\!\ud\xi\,\int\!\ud\xi'\,J(\xi,\theta)J(\xi',\theta')\,\frac1{\beta^2}\!\sum_{p_0,p'_0}G^{\down\up}(p_{(\theta,\xi)},k-p_{(\theta,\xi)},k-p_{(\theta',\xi')})\label{anglesusceptbcs}
\end{equation}
where $J(\xi,\theta)$ is the Jacobian, such that $1/V\sum_{\v p}\to\int\!\ud\theta\int\!\ud\xi J(\xi,\theta)$.

This susceptibility allows to analyze the superconducting instability without any prejudice on the form factor $f(\v p)$. The natural form factor is obtained by writing $\chi(\theta,\theta')$ in diagonalized form 
\begin{equation}
\chi(\theta,\theta')=\sum_n\chi_nf_n(\theta)f_n(\theta'),
\end{equation}
where $\chi_n$ and $f_n$ are, respectively, eigenvalues and eigenfunctions of the operator
\begin{equation}
f(\theta)\ \to\ \int\!\ud\theta'\,\chi(\theta,\theta') f(\theta').
\end{equation}
If $\chi_n$  diverges at an energy scale $\Lambda_c$, this indicates a (spin-, charge- or superconducting) instability with the form factor $f_n$. 

To simplify the calculations, one often selects a certain form factor $f(\v p)$ and calculates the susceptibility for $f_A(\v p)=f_B(\v p)=f(\v p)$.

\subsection{Density waves and flux phases}

The effect of a term creating electron-hole pairs is evaluated in the same way
as that of a two-particle source term. We consider operators of the form 
\begin{equation}
\Delta(\v r)=\frac12\sum_{\v r',\sigma}s_\sigma f(\v r'-\v r)c^\dagger_{\v r'\sigma}c^{}_{\v r\sigma},
\end{equation}
or in Fourier space
\begin{equation}
\Delta(\v q)=\frac12\sum_{\v p,\sigma}s_\sigma f(\v p)c^\dagger_{\v p\sigma}c^{}_{\v p+\v q\sigma}.
\end{equation}
The most simple cases are the magnetization ($s_\sigma=\sigma$) and the charge
($s_\sigma=1$), where $f(\v p)=1$, but many possibilities with non-trivial $\v
p$-dependencies exist. Two other examples (for $D=2$): if $f(\v p)$ is a function with $d_{x^2-y^2}$-symmetry,
$\v q=(\pi,\pi)$ and $s_\sigma=1$ ($\sigma$), then the nearest-neighbor-terms
in $\langle\Delta(\v q)\rangle$ yield circular charge (spin) currents flowing around the plaquettes of the square lattice with alternating directions (see 
Fig.~\ref{flux}). These charge and spin instabilities have been discussed a long time ago in the context of the excitonic insulator \cite{Halperin68}.

\begin{figure} 
        \centerline{\includegraphics[width=4cm]{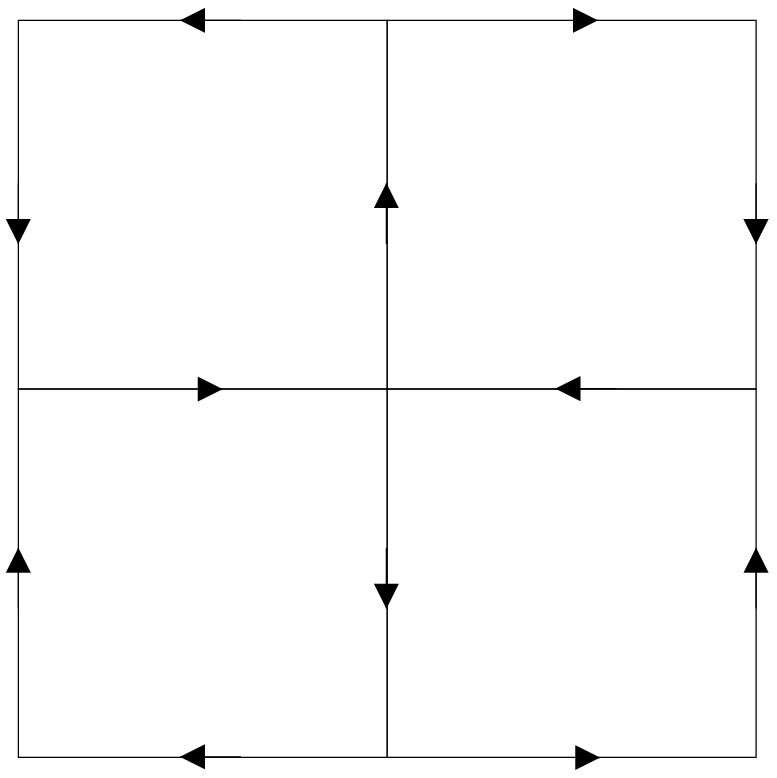}}
\caption{The pattern of charge (spin) currents along the bonds of 
the square lattice in a charge (spin) flux-phase.}\label{flux}
\end{figure}

The phase with circulating charge currents is actually identical to the
 flux-phase \cite{Affleck88,Kotliar88} and sometimes referred to as $d$-density
 wave, charge-current wave or orbital antiferromagnetism. Recently this order
 parameter was proposed to compete with $d$-wave
 superconductivity and to be responsible for the pseudo gap phase of the 
cuprates \cite{Chakravarty01}. 

The phase with circulating spin currents is called the spin flux-phase. Other names encountered in the literature are ``spin current wave'' or ``spin nematic state'' (because it is a state with broken rotational symmetry and unbroken time reversal symmetry). The low-temperature thermodynamics of both charge and spin flux-phases have been investigated in mean field approximation in Refs.~\cite{Nersesyan89,Nersesyan91}.

 We again considered two different form factors $f_A$ and
 $f_B$. This allows either to obtain an ``angle-resolved'' susceptibility
 $\chi^{s/c}(q,\theta,\theta')$ in analogy to Eq.~\eref{anglesusceptbcs}, or
 to select the form factor $f$ in advance and to choose $f_A=f_B=f$. The
 generalized charge- and spin susceptibilities are given by
\begin{eqnarray}
\chi^{s/c}_{BA}(q)&=&\frac1{V}\int_0^\beta\!\ud\tau\,e^{i\nu\tau}\left\langle\Delta^{}_B(\v q,\tau)\left(\Delta^\dagger_A(\v q)+\Delta^{}_A(-\v q)\right)\right\rangle\\
&=&\frac1{4V}\int_0^\beta\!\ud\tau\,e^{i\nu\tau}\sum_{\v p,\v p'}f_B(\v p)\left(\overline{f_A(\v p')}+f_A(\v p'+\v q)\right)\cdot\nonumber\\
& &\qquad\qquad\cdot\,\sum_{\sigma\sigma'}s_\sigma s_{\sigma'}\left\langle c^\dagger_{\v p\sigma}(\tau)c^{}_{\v p+\v q\sigma}(\tau)c^\dagger_{\v p'+\v q\sigma'}c^{}_{\v p'\sigma'}\right\rangle,
\end{eqnarray}
where  $q=(\nu,\v q)$. The most prominent case
 is $\v q=(\pi,\pi,\ldots)$, where $\Delta_{A/B}(\v q)$ can be made Hermitian if we choose $f(\v p+\v q)=\overline{f(\v p)}$. With this choice the spin- and charge susceptibilities can be written in perfect analogy with Eq.~\eref{chiBCS}
\begin{eqnarray}
\chi^{s/c}_{BA}(q)&=&\frac1{(\beta V)^2}\sum_{p,p'}f_B(\v p)\overline{f_A(\v p')}\left(G^{\up\up}(p',p+q,p)\mp G^{\up\down}(p',p+q,p)\right).\\
&=&\frac1{(\beta V)^2}\sum_{p,p'}f_B(\v p)\left[-\beta V\delta_{p,p'}+D^{ph}_q(p)\,\Gamma^{s/c}_q(p,p')\right]D^{ph}_q(p')\,\overline{f_A(\v p')},\label{chisc}
\end{eqnarray}
where $\Gamma^s_q(p,p')=-X\Gamma(p',p+q,p)$ and  $\Gamma^c_q(p,p')=(2-X)\Gamma(p',p+q,p)$. 

\subsection{Flow equations for the susceptibilities}

The RG formalism can be used to calculate the susceptibilities to leading
order in the infrared cutoff $\Lambda$. To see this, we rewrite Eqs.~\eref{chiBCS} and \eref{chisc} as
\begin{eqnarray}
\chi^{BCS}_{B\!A,\Lambda}(k)&=&\frac1{\beta V}\sum_p Z^{BCS}_{B,\Lambda,k}(p)D^{pp}_{\Lambda,k}(p)f_A(\v p)\label{chiZBCS}\\
\chi^{s/c}_{B\!A,\Lambda}(q)&=&-\frac1{\beta V}\sum_p Z^{s/c}_{B,\Lambda,q}(p)D^{ph}_{\Lambda,q}(p)f_A(\v p),\label{chiZsc}
\end{eqnarray}
where the effective field vertices $Z$ are defined as 
\begin{eqnarray}
Z^{BCS}_{B,\Lambda,k}(p)=f_B(\v p)+\frac1{\beta V}\sum_{p'}f_B(\v p')D^{pp}_{\Lambda,k}(p')\Gamma_{\Lambda,k}^{BCS}(p',p)\label{ZBCS}\\
Z^{s/c}_{B,\Lambda,q}(p)=f_B(\v p)-\frac1{\beta V}\sum_{p'}f_B(\v p')D^{ph}_{\Lambda,q}(p')\Gamma_{\Lambda,q}^{s/c}(p',p).\label{Zsc}
\end{eqnarray}
The vertex function $\Gamma^{BCS}$ in Eq.~\eref{ZBCS} can be expressed by an infinite series in the p-p irreducible part $I^{pp}$. Formally
\begin{eqnarray}
Z^{BCS}_B&=&f_B+f_BD^{pp}\Gamma^{BCS}\nonumber\\
&=&f_B+f_BD^{pp}I^{pp}+f_BD^{pp}I^{pp}D^{pp}I^{pp}+\cdots\nonumber\\
&=&f_B+Z^{BCS}_BD^{pp}I^{pp}\label{ZBCSequation}
\end{eqnarray}

To get the flow equation, we take the derivative with respect to $\Lambda$  of Eq.~\eref{ZBCSequation}. It gives three terms
\begin{equation}
\dot Z^{BCS}_B=\dot Z^{BCS}_BD^{pp}I^{pp}+Z^{BCS}_B\dot D^{pp}I^{pp}+Z^{BCS}_BD^{pp}\dot I^{pp}.\label{ZRG1}
\end{equation}
The last term of Eq.~\eref{ZRG1} is neglected for the same reason as in Eqs.~\eref{Rppderiv}. The remaining equation is iterated to give
\begin{eqnarray}
\dot Z^{BCS}_B&=&Z^{BCS}_B\dot D^{pp}I^{pp}+Z^{BCS}_B\dot D^{pp}I^{pp}D^{pp}I^{pp}+Z^{BCS}_B\dot D^{pp}I^{pp}D^{pp}I^{pp}D^{pp}I^{pp}+\cdots\nonumber\\
&=&Z^{BCS}_B\dot D^{pp}\,\Gamma^{BCS}\label{ZRG}
\end{eqnarray}
This is the RG equation for the field-vertex $Z^{BCS}_B$. The form factor $f_B$ enters as an initial condition $Z^{BCS}_{B,\Lambda_0,k}(p)\approx f_B(\v p)$.

It is now easy to obtain a RG equation for $\chi^{BCS}$ as well. From Eqs.~\eref{chiZBCS} and \eref{ZRG}
\begin{eqnarray}
\dot\chi^{BCS}&=&\dot Z^{BCS}_B D^{pp}f_A+ Z^{BCS}_B \dot D^{pp}f_A\nonumber\\
&=&Z^{BCS}_B\dot D^{pp}\,\Gamma^{BCS}D^{pp}f_A+ Z^{BCS}_B \dot D^{pp}f_A\nonumber\\
&=&Z^{BCS}_B\dot D^{pp}\left(f_A+\Gamma^{BCS}D^{pp}f_A\right)\nonumber\\
&=&Z^{BCS}_B\dot D^{pp}Z^{BCS}_A
\end{eqnarray}
In the last equation, we have introduced the function $Z^{BCS}_{A,\Lambda,k}(p)$, which is the same as $Z^{BCS}_{B,\Lambda,k}(p)$, except that the initial condition $f_B(\v p)$ is changed into $f_A(\v p)$.

Written out, the RG equations for the pairing susceptibility are
\begin{eqnarray}
\dot\chi^{BCS}_{B\!A,\Lambda}(k)&=&\frac1{\beta V}\sum_p Z^{BCS}_{B,\Lambda,k}(p)\dot D^{pp}_{\Lambda,k}(p)\,Z^{BCS}_{A,\Lambda,k}(p)\label{chiRGfinal}\\
\dot Z^{BCS}_{\diamond,\Lambda,k}(p)&=&\frac1{\beta V}\sum_{p'} Z^{BCS}_{\diamond,\Lambda,k}(p')\dot D^{pp}_{\Lambda,k}(p')\,\Gamma^{BCS}_{\Lambda,k}(p',p),\label{ZRGfinal}
\end{eqnarray}
where $\diamond=A$ or $B$, and $\Gamma^{BCS}_{\Lambda,k}(p',p)=\Gamma_\Lambda(p',k-p',k-p)$. Once the vertex function $\Gamma^{BCS}$ is known from the one-loop RG equation, one can solve the linear equation \eref{ZRGfinal} for the initial condition $Z^{BCS}_{\diamond,\Lambda_0,k}(p)=f_\diamond(\v p)$ and finally one can integrate Eq.~\eref{chiRGfinal} to obtain the susceptibility. 

All the steps can be repeated for the charge- and spin susceptibilities, just by replacing  the quantities $Z^{BCS}, \Gamma^{BCS},I^{pp}$ by  $Z^{s/c}, \Gamma^{s/c},I^{s/c}$ and the p-p propagator $D^{pp}$ by $-D^{ph}$. Thus, 
\begin{eqnarray}
\dot\chi^{s/c}_{B\!A,\Lambda}(q)&=&-\frac1{\beta V}\sum_p Z^{s/c}_{B,\Lambda,q}(p)\dot D^{ph}_{\Lambda,q}(p)\,Z^{s/c}_{A,\Lambda,q}(p)\label{chiRGfinalsc}\\
\dot Z^{s/c}_{\diamond,\Lambda,q}(p)&=&-\frac1{\beta V}\sum_{p'} Z^{s/c}_{\diamond,\Lambda,q}(p')\dot D^{ph}_{\Lambda,q}(p')\,\Gamma^{s/c}_{\Lambda,q}(p',p),\label{ZRGfinalsc}
\end{eqnarray}
where $\diamond=A$ or $B$, $\Gamma^{s}_{\Lambda,q}(p',p)=-X\Gamma_\Lambda(p',p+q,p)$ and $\Gamma^{c}_{\Lambda,q}(p',p)=(2-X)\Gamma_\Lambda(p',p+q,p)$. 

\section{Functional integral formulation}\label{grassmann}

In the functional integral formulation the annihilation- and creation operators $c^{}_{\v k\sigma}$ and $c^\dagger_{\v k\sigma}$ are replaced by anti-commuting Grassmann fields $\psi_{(\tau,\v k,\sigma)}$ and  $\bar\psi_{(\tau,\v k,\sigma)}$. They depend on  the ``imaginary time'' variable $\tau\in[0,\beta]$ in addition to the labels of the single-particle states, and they satisfy anti-periodic boundary conditions in the $\tau$ variable. 

The Hamilton operator is transformed into the action 
\begin{equation}
S[\psi]=-\int_0^\beta\ud\tau\left(\sum_{\v k\sigma} \bar\psi_{(\tau,\v k,\sigma)}(\partial_\tau-\mu)\psi_{(\tau,\v k,\sigma)}+H[\tau,\psi]\right),
\end{equation}
where $H[\tau,\psi]$ is obtained from the Hamiltonian by the replacements $c_{\v k\sigma}\to\psi_{(\tau,\v k,\sigma)}$ and  $c^\dagger_{\v k\sigma}\to\bar\psi_{(\tau,\v k,\sigma)}$. It is useful to take the Fourier transform 
\begin{equation}
\psi_{\sigma k}=\beta^{-1/2}\int_0^\beta\ud\tau\, e^{ik_0\tau}\psi_{(\tau,\v k,\sigma)},\label{Fourier}
\end{equation}
where $k=(k_0,\v k)$ contains the Matsubara frequency $k_0$. It allows to write the action as 
\begin{equation}
S[\psi]=S_0[\psi]-W[\psi],
\end{equation}
with
\begin{equation}
 S_0[\psi]=\sum_{\sigma,k}\bar\psi_{\sigma k}(ik_0-\xi_{\v k}) \psi_{\sigma k},
\end{equation}
where $\xi_{\v k}=e_{\v k}-\mu$ and
\begin{equation}
W[\psi]=\frac12\frac1{\beta V}\sum_{k_1,k_2,k_3}g(\v k_1,\v k_2,\v k_3) \sum_{\sigma, \sigma'}\bar\psi_{k_1\sigma} \bar\psi_{k_2\sigma'}\psi_{k_3\sigma'}\psi_{k_1+k_2-k_3\,\sigma}.\label{int}
\end{equation}
The partition function is given by the functional integral
\begin{equation} 
Z=\int{\cal D}\psi\, e^{S[\psi]},\label{Z}
\end{equation}
where ${\cal D}\psi$ is a short-hand notation for $\prod_{k\sigma}\ud\psi_{k\sigma}\ud\bar\psi_{k\sigma}$. The one- and two-particle Green's functions are given by
\begin{equation}
G(k)=-\langle\psi_{k\sigma}\bar\psi_{k\sigma}\rangle
\end{equation}
\begin{equation}
G^{\sigma\sigma'}(k_1,k_2,k_3)=\beta V\, \langle\psi_{k_1\sigma}\psi_{k_2\sigma'}\bar\psi_{k_3\sigma'}\bar\psi_{k_1+k_2-k_3\sigma}\rangle.\label{G2}
\end{equation}
where the average of a Grassmann expression $O[\psi]$ is defined as
\begin{equation}
\langle O[\psi]\rangle=\frac1Z\int{\cal D}\psi\, e^{S[\psi]}O[\psi].
\end{equation}
It is convenient to introduce the partition function with source term 
\begin{equation} 
Z[\eta]=\int d\mu_C[\psi]\,  e^{-W[\psi]+(\bar\eta,\psi)+(\bar\psi,\eta)},
\label{Zeta}
\end{equation}
where we used the short-hand notation 
 $(\bar\chi,\psi):=\sum_{\sigma k} \bar\chi_{\sigma k}\psi_{\sigma k}$ and the normalized
Gaussian measure is  defined by
\begin{equation} 
d\mu_C[\psi]:=\frac{{\cal D}\psi\ e^{(\bar\psi,C^{-1}\psi)}}{\int{\cal D}\psi\ e^{(\bar\psi,C^{-1}\psi)}}.\label{Gaussian}
\end{equation}
The connected part of the average of any Grassmann monomial is obtained as a functional derivative \cite{Negele}
\begin{equation}
\langle\psi_1\cdots\psi_n\bar\psi_{n+1}\cdots\bar\psi_{2n}\rangle_{\rm c}=
\left.\frac{\delta^{2n}\log Z[\eta]}{\delta\!\eta_{2n}\cdots
\delta\!\eta_{n+1}
\delta\!\bar\eta_{n}\cdots\delta\!\bar\eta_{1}}\right|_{\eta=0},
\end{equation}
where we have written $\psi_i$ instead of $\psi_{k_i\sigma_i}$.


\end{document}